\documentclass[fleqn,usenatbib]{mnras}

\usepackage{newtxtext,newtxmath}
\usepackage[T1]{fontenc}

\DeclareRobustCommand{\VAN}[3]{#2}
\let\VANthebibliography\thebibliography
\def\thebibliography{\DeclareRobustCommand{\VAN}[3]{##3}\VANthebibliography}

\usepackage{graphicx}	
\usepackage{amsmath}	
\usepackage{comment}
\usepackage{multirow}
\usepackage{soul}
\usepackage[dvipsnames]{xcolor}

\newcommand{\gadgetosaka}{{\sc GADGET4-Osaka}}
\newcommand{\Myr}{\ensuremath{\,\mathrm{Myr}}}
\newcommand{\pth}{P_{\mathrm{th}}}
\newcommand{\pram}{P_{\mathrm{ram}}}
\newcommand{\wsa}{S_{\mathrm{w}}}
\newcommand{\Husko}{H23}
\renewcommand{\paragraph}[1]{\textit{#1}. --}

\title[AGN jet evolution]{AGN jet evolution simulation with \textsc{GADGET4-OSAKA}}

\author[Dong, Wiliardy, Nagamine, et al.]{
Chenze Dong,$^{1,2}$\thanks{E-mail: dong-chenze@g.ecc.u-tokyo.ac.jp}
Abednego Wiliardy,$^{3}$
Kentaro Nagamine,$^{3,4,2,5,6}$
Yuri Oku,$^{7}$
Akira Mizuta,$^{8,9}$
\newauthor{Boon Kiat Oh,$^{10,11}$ Renyue Cen $^{7,12}$}
\\
$^{1}$Center for Data-Driven Discovery, Kavli IPMU (WPI), UTIAS, The University of Tokyo, Kashiwa, Chiba 277-8583, Japan\\
$^{2}$Kavli Institute for the Physics and Mathematics of the Universe, The University of Tokyo, 5-1-5 Kashiwanoha, Kashiwa, Chiba, 277-8583, Japan\\
$^{3}$Theoretical Astrophysics, Department of Earth and Space Science, Graduate School of Science, The University of Osaka, 1-1 Machikaneyama, Toyonaka, 
\\Osaka 560-0043, Japan\\
$^{4}$Theoretical Joint Research, Forefront Research Center, Graduate School of Science, The University of Osaka, Toyonaka, Osaka 560-0043, Japan\\
$^{5}$Department of Physics \& Astronomy, University of Nevada, Las Vegas, 4505 S. Maryland Pkwy, Las Vegas, NV 89154-4002, USA\\
$^{6}$Nevada Center for Astrophysics, University of Nevada, Las Vegas, 4505 S. Maryland Pkwy, Las Vegas, NV 89154-4002, USA\\
$^{7}$Center for Cosmology and Computational Astrophysics, Institute for Advanced Study in Physics, Hangzhou 310058, China\\
$^{8}$Astrophysical Big Bang Laboratory, RIKEN Pioneering Research Institute (PRI), Wako, Saitama, 351-0198, Japan\\
$^{9}$RIKEN Center for Interdisciplinary Theoretical and
Mathematical Sciences (iTHEMS), Wako, Saitama, 351-0198, Japan\\
$^{10}$Department of Physics, University of Connecticut, Storrs CT 06269, USA\\
$^{11}$Korea Institute for Advanced Study (KIAS), Seoul, South Korea\\
$^{12}$Institute of Astronomy, School of Physics, Zhejiang University, Hangzhou 310058, China.
}

\date{Accepted XXX. Received YYY; in original form ZZZ}

\pubyear{\the\year{}}

\begin{document}
\label{firstpage}
\pagerange{\pageref{firstpage}--\pageref{lastpage}}
\maketitle


\begin{abstract}
Active galactic nuclei (AGN) jets are powerful drivers of galaxy evolution, depositing energy and momentum into the circumgalactic and intracluster medium (CGM/ICM) and regulating gas cooling and star formation. 
We investigate the dynamics of jet evolution in the self-similar regime using the smoothed particle hydrodynamics (SPH) code \gadgetosaka{}, systematically vary jet-launching schemes, artificial-viscosity prescriptions, mass resolution, and jet lifetimes and compare the results with grid-based simulation. 
Our analysis combines quantitative diagnostics of jet size and energetics with detailed morphological and thermodynamic characterizations from slice maps and phase diagrams. 
We find that jet lobe growth follows analytic self-similar scaling relations and converges with resolution, but is highly sensitive to the choice of artificial viscosity. 
While the overall jet size tracks self-similar predictions, the partitioning of thermal and kinetic energy departs significantly from the idealized picture, reflecting enhanced dissipation and mixing, which is consistent with the jet propagation in grid-based simulations.
These results establish robust benchmarks for SPH-based jet modeling, provide insight into the physical and numerical factors shaping jet--medium interactions, and lay the groundwork for future studies of AGN feedback in realistic galactic and cluster environments.
\end{abstract}

\begin{keywords}
galaxies: jets --- galaxies: evolution --- galaxies: active --- galaxies: nuclei --- methods: numerical --- hydrodynamics
\end{keywords}



\section{Introduction}
\label{sec:intro}

Active galactic nuclei (AGN) are widely recognized as key drivers of galaxy evolution.  
By injecting vast amounts of energy and momentum into their surroundings, AGNs regulate star formation and shape the thermodynamic state of the circumgalactic and intergalactic media (CGM/IGM).  
AGN feedback is typically categorized into two complementary modes (see, e.g., \citealt{Merloni:2008, Heckman:2014, Dubois:2016}).  
In the radiative (quasar) mode, rapid black-hole accretion powers luminous, quasi-isotropic winds \citep{Feruglio:2015, Tombesi:2015}. 
In the kinetic (maintenance or radio) mode, more modest accretion launches highly collimated jets that prevent hot gas from cooling and accreting onto the galaxy \citep{Tortora:2009}. 
Together, these feedback channels establish the observed black hole--galaxy scaling relations and shape the star-formation histories of galaxies \citep{Bower2006,Croton2006}.

AGN jets are among the most powerful astrophysical phenomena, capable of impacting not only their host galaxies but also the surrounding CGM, intracluster (ICM), and even the IGM. These jets, composed of highly energetic particles --- primarily electrons and protons --- have been extensively studied through multi-wavelength observations \citep{Croston:2018, Turner:2018}. X-ray imaging has revealed vast cavities in galaxy clusters (see, e.g., \citealt{Birzan:2004, McNamara:2005, Wise:2007, Hlavacek-Larrondo:2012,Shin:2016, Liu:2019}), interpreted as relics of jet-driven outflows, while radio observations trace synchrotron emission from relativistic particles accelerated within the jets (\citealt{Burke-Spolaor:2017, Gendron-Marsolais:2017, DiGennaro:2018}). Additional indirect signatures, such as X-ray-bright rims around radio lobes (\citealt{Fabian:2000}; see also \citealt{Prunier:2025} for simulation comparison), reinforce the view that AGN jets can dramatically reshape their environment on both galactic and extragalactic scales.

As AGN jets propagate through the ambient medium, they interact dynamically with the ambient gas, altering the physical properties of both the jets and the surrounding medium. These interactions play a key role in the redistribution of energy, leading to the heating of the CGM/IGM \citep[e.g.,][]{Yamada:2001, McNamara:2007, Dong:2023} and driving large-scale outflows \citep[]{Lister:2005,Guillard:2012,Morganti:2015,Venturi:2021}. Such processes are central to galaxy evolution, especially in massive elliptical galaxies, where they suppress star formation by inhibiting gas cooling \citep{McNamara:2006, Guillard:2015}. Over cosmic times, this feedback mechanism has had a profound impact on the growth of galaxies and the assembly of large-scale structures. In addition to the galaxy cluster study, the AGN jet also provides a plausible mechanism for the formation of the Fermi bubble around the Milky Way \citep{Cheng:2011, Guo:2012, Yang:2012}. 

Rapid advancements in cosmological simulations have led to increasingly sophisticated methodologies for modeling the large-scale effects of radio-mode AGN feedback.
Early sub-grid models typically employed isotropic “thermal bombs” or broad-cone momentum kicks \citep[e.g.,][]{DiMatteo05, Springel05a, Dubois13}.  While effective at quenching star formation, these approaches failed to reproduce the elongated X-ray cavities and narrow radio lobes observed.   
More recent efforts have introduced collimated kinetic jets, with parameters such as opening angle, duty cycle, and mechanical efficiency calibrated either from small-scale magneto-hydrodynamic (MHD) simulations or from empirical scaling relations (e.g., \citealp{Weinberger:2017, Bourne2017, Cielo2018, Dave:2019, Perucho2019}).  
These studies demonstrate that the details of the jet-injection scheme --- including mass loading, nozzle geometry, artificial viscosity, and numerical resolution --- strongly influence lobe length, morphology of bow shock, and level of metal mixing.

A systematic comparison of these numerical ingredients remains lacking. 
Notablely, \citet{Husko2023a} utilizing the smoothed-particle hydrodynamics (SPH) code \textsc{SWIFT} has demonstrated that self-similar jet growth can be reproduced when physical parameters align with theoretical predictions.
Their work was conducted under the following specifications: 
(i) the use of a pre-allocated reservoir of jet particles,
(ii) a time-dependent artificial viscosity (AV) model, and
(iii) ignoring all physics except hydrodynamics. 
However, it remains unclear whether these findings can be extended to other modern SPH formulations when the treatment of dissipation and kernel bias varies.  
Furthermore, the sensitivity of jet lobe evolution to factors such as jet lifetime and gas mass resolution remains only partially understood, despite their relevance to interpreting radio and X-ray observations of relic AGN bubbles.

In this paper, we follow the reservoir-based kinetic-jet setup of \citet{Husko2023a} (hereafter \Husko{}) and implement the method in another modern SPH code \gadgetosaka{}.  
Our objectives are threefold: \\
1) Demonstrate the numerical reproducibility of self-similar lobe growth under controlled conditions; \\
2) Quantify the sensitivity of jet evolution to key numerical and physical setups; \\
3) Provide benchmark simulations for future model validation of jet feedback implementations.

In this work, we adopt the homogeneous circumgalactic configuration in \Husko{} and a similar particle-reservoir setup for jet launch.
In addition to variations in mass resolution, we investigate how viscosity models, jet reservoir arrangement, and jet active time affect jet propagation. 

The remainder of the paper is structured as follows.
Section~\ref{sec:analytical-model} reviews the self-similar analytic model that underpins our interpretation of lobe growth.
Section~\ref{sec:numerical-implementation} describes the implementation of \gadgetosaka{}, including the jet injection algorithm and the diagnostics used to identify jet lobes, and bow shocks.
Section~\ref{sec:results} presents our results, beginning with the fiducial simulation and following comparisons across all parameter variations.
In Section~\ref{sec:discussion}, we place these findings in the broader context of previous numerical and observational studies.  
Finally, Section~\ref{sec:conclusions} summarizes our main conclusions and discusses future directions, including the incorporation of additional physics.


\section{Analytical model}
\label{sec:analytical-model}

\begin{figure}
\centering
\includegraphics[width=\columnwidth]{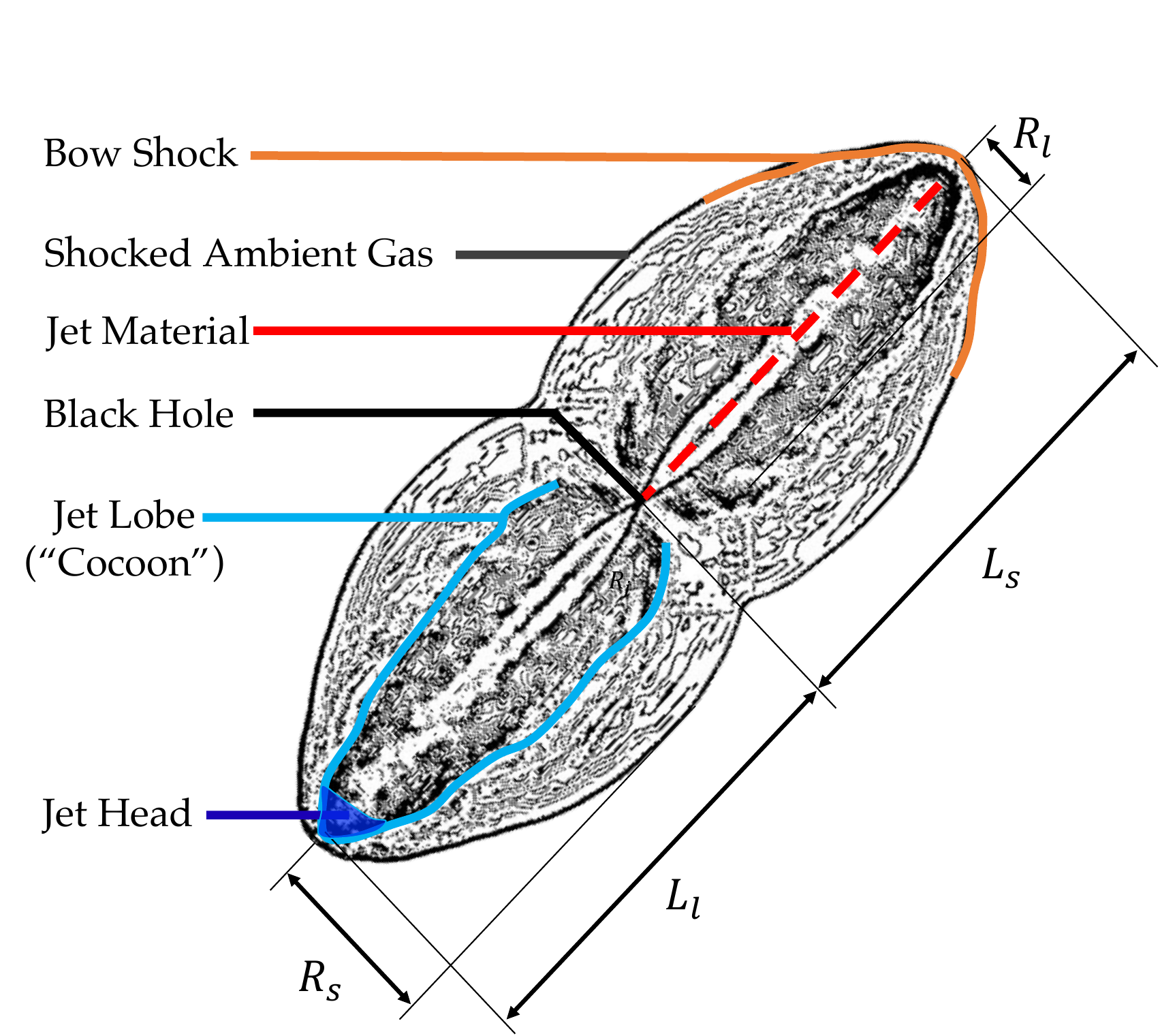}
\caption{Schematic representation of the jet structure derived from the slice plot of our fiducial simulation. The structure comprises a central jet , which terminates at the jet head and inflates a surrounding lobe of shocked jet material. 
The lobe (also known as "cocoon") with length $L_\ell$ and radius $R_\ell$} is embedded within a larger region of shocked ambient gas, bounded by a bow shock that defines the overall length $L_s$  and radius $R_s$ bow shock. This diagram illustrates the key components used in our morphological and dynamical analysis.
\label{fig:jet-schematic}
\end{figure}

The jet structure derived from our fiducial simulation is illustrated in Figure~\ref{fig:jet-schematic}. Following an outside-in order, the jet structure can be broadly decomposed into three primary components: the bow shock, the jet lobe, and the jet material. 
The bow shock predominantly consists of shocked ambient gas that has been compressed as the jet propagates, resulting in an elevated density and temperature compared to the unperturbed background. 
The outer boundary of the shocked ambient gas is marked by a bow shock that separates this shocked region from the undisturbed ambient medium.
Within the bow shock, the jet lobe manifests as a cavity of low-density, high-temperature gas resulting from the interaction of the jet with the ambient medium. 
This region consists of a turbulent mixture of shocked jet material and entrained ambient gas.
The tip of the jet lobe, which is known as the `jet head', is the site of jet material mixing with the ambient gas. 
The jet (material) itself remains collimated along the symmetry axis, terminating at a strong shock near the jet head, where much of its kinetic energy is transferred to the lobe and bow shock.

It is essential to compare our simulated jet lobes with analytical models of self-similar jet evolution \citep[e.g.,][]{Falle1991, Kaiser1997, Komissarov1998, Begelman1989, Bromberg2011}.
In this framework, the jet is launched with a specific half-opening angle, $\theta_\mathrm{j}$, and is characterized by key physical parameters including the jet power ($P_\mathrm{j}$), launch velocity ($v_\mathrm{j}$), and the density ($\rho$), pressure ($p$), and temperature ($T$) of the ambient medium. 
For a purely kinetic (non-relativistic) jet, the mass-loading rate is determined by the balance between kinetic energy flux and power, yielding
\begin{equation}
    \dot{M}_\mathrm{j} = 2 \frac{P_\mathrm{j}}{v_\mathrm{j}^2}.
\end{equation}
Here, we neglect relativistic corrections given the galactic scale of simulation and assume symmetry between the two bipolar jets, so that the quoted parameters describe the combined effect of both lobes.


The two characteristic length scales governing the jet propagation \citep[][]{Falle1991,Kaiser1997,Komissarov1998} are given by
\begin{equation}
    L_1 = \frac{1}{\theta_\mathrm{j}} \sqrt{\frac{2}{\pi \rho} \sqrt{\frac{\dot{M}_j^3}{2 P_\mathrm{j}}}} = \frac{2}{\theta_\mathrm{j}} \sqrt{\frac{P_\mathrm{j}}{\pi \rho v_\mathrm{j}^3},}
    \label{eq:L1}
\end{equation}
\begin{equation}
\label{eq:L2}
    L_2 = \left( \frac{P_\mathrm{j}^2 \rho}{p^3} \right)^{1/4}.
\end{equation}

Here, $L_1$ marks the distance at which the mass swept up by the jet becomes comparable to the mass injected by the jet itself. In the early stages, when the jet length satisfies $L_\mathrm{j} \ll L_1$, the jet remains much denser than the ambient medium and thus propagates nearly unimpeded. This phase, referred to as `ballistic' in \Husko{}, is characterized by linear growth of the jet length with time, $L_\mathrm{j} = v_\mathrm{j} t$. 

The scale $L_1$ sets the transition from this freely expanding regime to one in which the internal jet pressure begins to balance the pressure of the surrounding shocked gas, leading to partial collimation. 
In contrast, $L_2$ marks the point at which the internal pressure of the lobe approaches equilibrium with the ambient pressure. 
Once the jet extends beyond $L_2$, its expansion slows as the ambient pressure becomes dynamically significant and the advance speed of the bow shock approaches the sound speed of the external medium \citep{Komissarov1998, Husko2023a}.
Together, these two characteristic scales delineate whether the system evolves in a momentum-driven or pressure-confined regime, and provide essential benchmarks for interpreting the behavior of AGN jets in both simulations and observations.

In the intermediate regime defined by $L_1 \ll L_\mathrm{j} \ll L_2$, the jet evolution is expected to follow a self-similar solution. 
In this phase, both the external pressure and the swept-up mass flux are dynamically negligible, leaving jet power and ambient gas density as the only relevant parameters. 
Since no intrinsic length or time scale can be derived from these two quantities alone, dimensional analysis implies that the system evolves self-similarly \citep{Sedov1959}.

In a uniform medium, the jet length grows as the density of the jets increases: 
\begin{equation}
\label{Eq:length-evo-self-similar}
    L_\mathrm{j}(t) = \left(\frac{P_\mathrm{j} t^3}{\rho} \right)^{1/5}.
\end{equation}


For a more general ambient medium following a power-law density profile, $\rho(r) = \rho_0 (r / r_0)^{-\beta}$, the self-similar jet length evolves as \citet{Kaiser2007}
\begin{equation}
\label{Eq:length-evo-self-similar-final}
    L_\mathrm{j}(t) = c_1 \left(\frac{P_\mathrm{j} t^3}{\rho_0 r_0^\beta} \right)^{1/(5-\beta)},
\end{equation}
where $c_1$ is a dimensionless constant defined by
\begin{equation}
\label{Eq:c1}
    c_1 = \left(\frac{A^4}{18 \pi} \frac{(\gamma_A + 1) (\gamma_L - 1) (5 - \beta)^3}{9 \left(\gamma_L + \frac{A^2}{2}(\gamma_L - 1) \right) - 4 - \beta}   \right)^{1/(5-\beta)}.
\end{equation}
Here, $A=R_\ell/L_\ell$ is the aspect radio of lobe; $R_\ell$ represents the radius of the lobe, while $L_\ell\approx L_J$ is the length of the lobe. 
$\gamma_A$ and $\gamma_L$ are the adiabatic indices of the ambient gas and the lobe, respectively. In this work, we adopt $\gamma_A = \gamma_L = 5/3$, as we neglect relativistic effects.

Taking the time derivative of Equation~\eqref{Eq:length-evo-self-similar-final} yields the self-similar advance speed of the lobe head:
\begin{equation}
\label{Eq:self-similar-vel}
    v_\mathrm{H}(t) = \frac{3 c_1}{5-\beta} \left(\frac{P_\mathrm{j}}{\rho_0 r_0^\beta} \right)^{1/(5-\beta)} t^{\frac{\beta-2}{5-\beta}}.
\end{equation}
This framework provides a theoretical baseline against which we can compare our simulated jets to assess the degree of self-similarity and the influence of numerical effects.

\begin{table*}
\centering
 \caption{Summary of simulations in this study.}
 \label{table:list-of-simulation}
 \scalebox{1.}{
\begin{tabular}{cccccc}
\hline
\multicolumn{6}{c}{\textbf{Fixed Parameters for all runs}} \\ \hline
\multicolumn{3}{c|}{Jet Setup} & \multicolumn{3}{c}{Simulation Configuration} \\ \hline
$v_{\mathrm{jet}} $ & \multicolumn{2}{c|}{$15000\,\mathrm{km\, s^{-1}}$} & $\rho_{\mathrm{0}} $ & \multicolumn{2}{c}{$1.2\times10^{-26}\,\mathrm{g\,cm^{-3}}$} \\
$P_{\mathrm{jet}} $ & \multicolumn{2}{c|}{$10^{46}\,\mathrm{erg/s}$} & $T_{\mathrm{0}}$ & \multicolumn{2}{c}{$10^{7.2}\,\mathrm{K}$} \\
$\theta_{\mathrm{jet}}$ & \multicolumn{2}{c|}{$10\,\deg$} & $P_{\mathrm{0}}$ & \multicolumn{2}{c}{$2.66\times10^{-11}\,\mathrm{g/(cm\cdot s^2)}$} \\
\multicolumn{1}{l}{} & \multicolumn{1}{l}{} & \multicolumn{1}{l|}{} & $K_{\mathrm{0}}$ & \multicolumn{2}{c}{$4.24 \times 10^{32}\,\mathrm{cm^4/(g^{2/3}\cdot s^2)}$} \\
 & \multicolumn{2}{c|}{} & $(L_x, L_y, L_z)$ & \multicolumn{2}{c}{$(300, 300, 1200)\,\mathrm{kpc}$} \\
 & \multicolumn{2}{c|}{} & $t_\mathrm{sim}$ & \multicolumn{2}{c}{$100\Myr$} \\ \hline
\textbf{Varied Parameters} &  &  &  &  &  \\ \hline
Run Name & $t_\mathrm{jet} [\mathrm{Myr}]$ & $m_{\mathrm{gas}} [10^5 M_\odot]$ & $L_{\mathrm{sep}} [\mathrm{kpc}]$ & AV Model & Launching Scheme \\ \hline
FID & 100 & 1.81 & 1 & TDV, $\alpha_{\mathrm{max}}=2$ & RG \\
\multicolumn{6}{c}{Artificial Viscosity Variants} \\ \hline
StV & 100 & 1.81 & 1 & StV, $\alpha=2$ & RG \\
HVisc & 100 & 1.81 & 1 & TDV, $\alpha_{\mathrm{max}}=5$ & RG \\
LLVisc & 100 & 1.81 & 1 & TDV, $\alpha_{\mathrm{max}}=0.5$ & RG \\
\multicolumn{6}{c}{Launching Scheme Variants} \\ \hline
RR & 100 & 1.81 & 1 & TDV, $\alpha_{\mathrm{max}}=2$ & RR \\
SR & 100 & 1.81 & 1 & TDV, $\alpha_{\mathrm{max}}=2$ & SR \\
\multicolumn{6}{c}{Mass Resolution Variants} \\ \hline
Hi & 100 & 0.93 & 0.8 & TDV, $\alpha_{\mathrm{max}}=2$ & RG \\
LoA & 100 & 6.11 & 1.5 & TDV, $\alpha_{\mathrm{max}}=2$ & RG \\
LoB & 100 & 14.48 & 2 & TDV, $\alpha_{\mathrm{max}}=2$ & RG \\
LoC & 100 & 48.87 & 3 & TDV, $\alpha_{\mathrm{max}}=2$ & RG \\
LoD & 100 & 115.84 & 4 & TDV, $\alpha_{\mathrm{max}}=2$ & RG \\
\multicolumn{6}{c}{Jet Time Variants} \\ \hline
T50 & 50 & 1.81 & 1 & TDV, $\alpha_{\mathrm{max}}=2$ & RG \\
T20 & 20 & 1.81 & 1 & TDV, $\alpha_{\mathrm{max}}=2$ & RG \\
\multicolumn{6}{c}{Grid-Based Variants}\\ \hline
MESH & \multicolumn{5}{c}{See Section \ref{subsec:sim_var}}

\end{tabular}
 
}
 
\end{table*}
\section{Numerical method}
\label{sec:numerical-implementation}

In this study, we employ the \gadgetosaka{} code \citep{Romano:2022a, Romano:2022b, Oku:2022, Oku:2024}, a form of the SPH code GADGET-4 \citep{Springel:2021}, to perform our hydrodynamical simulations. 
Following the methodology of \Husko{}, we include only hydrodynamic interactions among SPH particles and deliberately omit additional physics such as gravity, radiative cooling, magnetic fields, and cosmic rays. 
This stripped-down setup enables a clean examination of jet propagation in an idealized, homogeneous medium, thereby allowing direct comparison with analytic models and isolating numerical effects from physical complexities.

We describe the SPH methodology implemented in \gadgetosaka{} in Section~\ref{subsec:SPH_treatment}, followed by an overview of the simulation setup in Section~\ref{subsec:setup}. 
Variations in jet launch schemes are discussed in Section~\ref{subsec:jet_launch}, and the full suite of feedback variants explored in this study is summarized in Section~\ref{subsec:sim_var}. 
We also introduce comparisons with grid-based simulations by \citet{Mizuta2006,Mizuta2026} in Sec~\ref{subsec:sim_var}.
Finally, the diagnostics used to quantify jet propagation are detailed in Section~\ref{subsec:measurements}.

\subsection{Numerical Schemes} \label{subsec:SPH_treatment}

\subsubsection{SPH kernel formulation}

We adopt the Wendland C4 kernel \citep{Dehnen:2012} with bias correction for all SPH interactions, setting the number of neighbors per particle to $n_\mathrm{ngb} = 200$. This choice offers a good compromise between stability and resolution, while suppressing pairing instability.

Our simulations employ the pressure-entropy formulation of SPH introduced by \citet{Hopkins2013a}, which offers significant advantages in handling contact discontinuities and shock fronts. 
The pressure-based formulation is known to better capture fluid mixing and to mitigate numerical surface-tension effects that can artificially suppress Kelvin-Helmholtz and Rayleigh-Taylor instabilities. 
Therefore, such a formulation becomes the default choice in the Osaka feedback models \citep{Oku:2024}. 
This choice contrasts with \Husko{}, which uses an energy--density formulation, whereas \gadgetosaka{} does not. 

Although gravitational effects are not incorporated within our simulations, we still adopted the Jeans pressure floor following the prescription of \citet{Hopkins:2011} and \citet{Kim:2016}.
Such a pressure floor is useful for maintaining numerical stability in the scenario of jet simulations characterized by highly supersonic flow dynamics.

\subsubsection{Artificial Conductivity and Artificial Viscosity}

\gadgetosaka{} incorporates both artificial conductivity and artificial viscosity. 
Artificial conductivity is crucial for mitigating density and temperature disparities that occur at shock interfaces, which is included with the velocity-based conductivity signal speed \citep{wadsley2008}.
To enhance the resolution of shock waves, a time-dependent limiter \citep{Borrow2022} is applied to the artificial conduction in these regions.

Artificial viscosity (AV) is critical for resolving shocks and handling velocity discontinuities. In this study, we examine two AV formulations: the standard viscosity (StV) and time-dependent viscosity (TDV) schemes. 
Both are implemented in \textsc{Gadget-4} \citep{Springel:2021}. 
The standard viscosity has a constant viscosity coefficient for all SPH particles throughout the simulation, which is intended to capture the shocks around the supersonic flows like the AGN jet. 
The TDV formulation implemented with the Cullen--Dehnen switch \citep{Cullen2010, Hu:2014}, by contrast, dynamically adjusts the viscosity coefficient on the basis of local flow conditions, reducing the numerical dissipation away from shocks while preserving the accuracy in supersonic regions.

To further improve the performance of AV in shock-capturing, we apply the higher-order velocity gradient estimator proposed by \citet{Hu:2014} and get the particle relative velocity from velocities linearly reconstructed at the particle midpoint \citep{frontiere2017,rosswog2020} with the Balsara slope limiter \citep{balsara2004} and the minmod limiter, which sharpens discontinuities and minimizes spurious angular momentum transport.
Additionally, we adopt the Balsara switch \citep{Balsara:1995} to suppress AV in shear-dominated flows, preventing excessive damping of turbulence and vorticity.

\subsubsection{Time Integration}
The hydrodynamical force is integrated using the second-order predictor-corrector method \citep{Springel:2021} with the timestep limiter \citep{saitoh2009,durier2012,Oku:2024}. This ensures proper adjustment of the particle timestep during jet propagation. We adopted a Courant number of $0.3$ in all our simulations.

\subsection{Simulation Setup} 
\label{subsec:setup}

We follow the initial condition setup by \Husko{}. 
The simulation is performed in a cuboid periodic simulation box with a side length of $(L_x, L_y, L_z) = (300, 300, 1200) \,\mathrm{kpc}$, where the $z$ axis is extended to accommodate the direction of jet propagation. We emphasize to the readers that the periodic configuration is implemented solely for facilitating the accounting of the energy budget; the jet structures, however, do not traverse the periodic boundary upon the end of the simulation.
The total simulation time is set to $T_{\mathrm{sim}} = 97.8\Myr \approx 100\Myr$.

To initialize a uniform and static ambient medium, all gas particles are assigned a constant density of $\rho_{0} = 1.2 \times 10^{-26}\,\mathrm{g/cm^3}$ and a temperature $T_{0} = 10^{7.2}\,\mathrm{K}$.
The particles are placed on a regular grid and given zero initial velocity to ensure a static condition. 
The interparticle spacing is determined by the mass resolution. 
In our fiducial run (FID), the spacing is $L_{\mathrm{sep}} = 1\,\mathrm{kpc}$, corresponding to a gas particle mass of $m_{\mathrm{gas}} = 1.81 \times 10^5\,\mathrm{M_\odot}$.
In the resolution-variant runs (Hi, LoA, LoB, LoC, LoD; see \ref{table:list-of-simulation}), this spacing is adjusted according to the respective mass resolution to explore the effects of resolution on jet propagation.

\subsection{Jet launching mechanisms} \label{subsec:jet_launch}

A key numerical challenge in modeling AGN jet feedback using SPH-based codes arises when injecting energy at a constant power. Continuous energy injection eventually depletes the gas supply in the vicinity of the black hole, leading to the formation of an unphysical low-density cavity around the jet origin \citep{Vogelsberger:2013, Barai:2014, Barai:2016, Weinberger:2017}. This artifact stems from the repeated consumption of local gas particles, prompting the black hole to increasingly draw from more distant neighbors, thereby disrupting the local gas structure.

This issue becomes particularly pronounced when the jet operates on short time steps, accelerating the depletion process. Additionally, if launched particles are not properly tagged or tracked, there is a risk that the same particle may be “kicked” multiple times, further exacerbating inaccuracies in the simulation of jet propagation and energy deposition.

In this study, we adopt a reservoir-based jet launching scheme in which a pre-allocated set of gas particles is designated at the initial conditions for subsequent injection. The total number of reservoir particles is determined by total energy of the jet $E_\mathrm{jet} = P_\mathrm{jet}T_\mathrm{jet}$ and the energy of one jet launching event $E_{\mathrm{launch}}=m_{\mathrm{gas}}v_\mathrm{jet}^2/2$. In our fiducial run, we set the jet-launching duration equal to the simulation time, $T_\mathrm{jet} = T_\mathrm{sim} = 97.8 \Myr \approx 100\Myr$. 
For comparison, we also examine shorter-duration scenarios with $T_\mathrm{jet} = 0.5 T_\mathrm{sim} \approx 50\,\Myr$ (T50) and $T_\mathrm{jet} = 0.2 T_\mathrm{sim} \approx 20\,\Myr$ (T20), in which the jet is active only during the early phase of the simulation.

This reservoir method, originally proposed by \Husko{}, offers a computationally efficient alternative to on-the-fly particle selection by avoiding repeated neighbor searches. It is particularly well-suited to our goal of studying jet propagation in a controlled environment. However, in more realistic simulations of self-regulated black hole accretion, reservoir particles would need to be spawned dynamically in the vicinity of the black hole, which would require additional implementation beyond our current \gadgetosaka{} setup.

In this idealized simulation, the AGN jet is launched at a fixed power $P_\mathrm{j} = 10^{46}\,\mathrm{erg/s}$.
Given our choice of a constant launching velocity $v_\mathrm{j}=15000\,\mathrm{km/s}$, 
the time interval between successive jet events, i.e., the injection of a pair of gas particles with mass $m_\mathrm{gas}$, is given by 
\begin{equation}
    \mathrm{d}t_\mathrm{j} = \frac{m_\mathrm{gas} v_\mathrm{j}^2}{P_\mathrm{j}}.
\end{equation}
This expression represents the time required for the AGN to accumulate sufficient kinetic energy to launch one jet event. As a result, $\mathrm{d}t_\mathrm{j}$ depends directly on the mass resolution of the simulation.
The initial temperature and density of the launched gas are set equal to those of the ambient medium, ensuring consistent thermodynamic conditions at injection.

\begin{figure}
\centering
\includegraphics[width=\columnwidth]{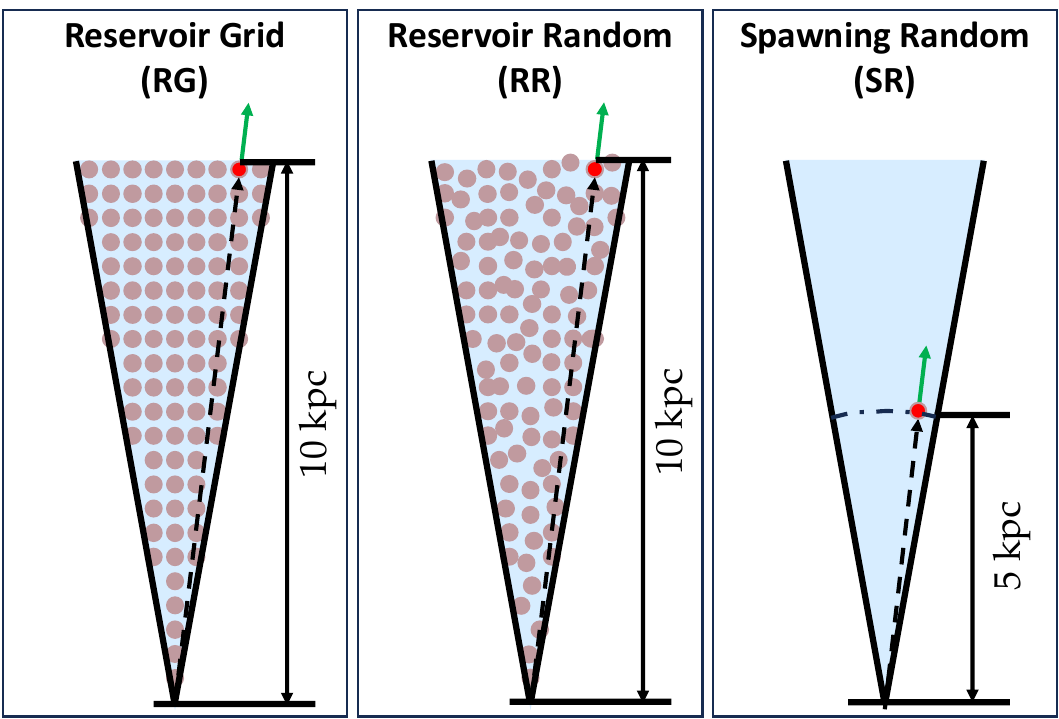}
\caption{
Schematic illustration of the jet launching mechanism. The red particle represents a newly injected SPH particle within the simulation box. The pink circles denote reservoir particles that will be launched at later stages of the simulation. The green arrow indicates the direction of injected jet particle, which is defined by the dashed arrow connecting the origin to the newly injected SPH particle.}
\label{fig:jet-launching-machanism}
\end{figure}

We find that the spatial configuration of the reservoir particles can also influence jet morphology and propagation. To explore this, we implement three different particle placement schemes:
reservoir grid (RG),  reservoir random (RR) and Spawning Random (SR). 
We illustrate the launching schemes in Figure ~\ref{fig:jet-launching-machanism}.

\textbf{Reservoir Grid (RG)} --- 
In this scheme, the reservoir particles are placed on a Cartesian lattice and then trimmed into a biconical volume defined by the jet’s half-opening angle, $\theta_{\mathrm j}$, out to a maximum distance along the $z$-axis.  Particles are injected in an `outside-in' order, so the first particles launched lie closest to the bow shock and interact immediately with the ambient medium.  Within each radial layer, the launch order is randomised to avoid artificial symmetry.
We stress that `grid' refers only to the initial spatial arrangement of reservoir particles; the hydrodynamics remain fully Lagrangian, and no fixed mesh is introduced in the simulation.

\textbf{Reservoir Random (RR)} --- 
Like the RG scheme, the RR model relies on a pre-allocated set of particles, but their initial positions are chosen at random inside the biconical launch volume.  This stochastic placement yields a more isotropic sampling of directions, thereby avoiding the drop in particle density toward the axis inherent to a trimmed grid.

In contrast, the trimmed-grid approach of RG reduces the number of particles per layer as one moves toward the center, resulting in a less uniform distribution within the specified opening angle. While randomization in the RR model can enhance spatial uniformity, it also introduces the risk of particles being positioned too closely together, which may affect the simulation's efficiency and duration.

\textbf{Spawning Random (SR)} --- 
In this scheme, jet particles are `spawned' on the fly rather than drawn from a pre-defined reservoir.  At each jet event, we inject one particle pair in opposite directions; the launch vector is chosen randomly within the prescribed half-opening angle, and the injection point is fixed at a distance of 5 kpc from the black-hole position.

A spawning approach is particularly attractive for large cosmological runs that track many black holes, whose jet duty cycles and particle budgets are not known a priori.  Nevertheless, the \textit{spawned} SPH particles in this setup are still pre-allocated in memory because of the technical challenges associated with introducing new gas particles in \textsc{Gadget-4}. 
We leave the integration of our jet-launching scheme into cosmological simulations to future work, which involves collecting and reusing the memory of removed particles.

All three launching schemes introduce additional mass into the simulation \footnote{Note that, for all three schemes, jet particles are assigned to the computer memory ahead of time, but are not considered in the force calculation before they are launched. 
}, set by the jet’s mass-loading rate, 
\begin{equation}
    \dot{M}_\mathrm{j} = 2 \ \frac{P_\mathrm{j}}{v_\mathrm{j}^2}.
\end{equation}
In the fiducial run, this yields $\dot{M}_\mathrm{j} = 141 M_\odot / \mathrm{yr}$, or $1.41 \times 10^{10} M_\odot$ over the 100 Myr simulation, which is less than $0.1$ per cent of the total ambient gas mass in the initial condition. 
Because self-gravity is disabled in these idealised runs, such a modest mass surplus has a negligible dynamical impact.

Prior to launch, reservoir particles are `frozen': they neither interact hydrodynamically nor appear in neighbour lists of active gas particles.  The total number of frozen particles equals the expected number of jet events during the prescribed jet-active time, $t_{\rm jet}$.  
Each particle is activated only when its turn to be launched is reached, at which point it acquires the jet velocity and full hydrodynamic coupling.

\subsection{Simulation Variants} 
\label{subsec:sim_var}

To test the robustness and flexibility of our model, we conducted 11 simulations in addition to the fiducial run; their settings are summarised in Table \ref{table:list-of-simulation}.  Each variant isolates one numerical or physical ingredient:
\begin{itemize}
\item Artificial-viscosity variants (StV, HVisc, LLVisc) --- The StV employs a standard constant-viscosity model to evaluate the effect of viscosity on jet propagation. 
The HVisc, and LLVisc runs are still based on the time-dependent viscosity model, but we varied the max allowed viscosity coefficient ($\alpha_{\mathrm{max}}=5, 0.5 $) to investigate the impact of this parameter.

\item Launching-scheme variants (RR, SR) --- alter the way jet particles are introduced as shown in Section \ref{subsec:jet_launch}: RR randomizes the order in which reservoir particles are injected, whereas SR spawns new particles on-the-fly, sampling directions stochastically.  These runs test the sensitivity of jet morphology to the injection geometry.
\item Mass-resolution series (Hi, LoA--LoD) --- spans nearly two orders of magnitude in particle mass, with each step differing by a factor of $\sim 3$, allowing us to evaluate how resolution affects lobe growth and bow shock structure in cosmological simulation contexts.
\item Jet-lifetime variants (T50, T20) --- truncate the period of active injection to approximately 50 Myr and 20 Myr, respectively, in order to track the evolution of `fossil' lobes once the power source is switched off.
\end{itemize}

Lastly, we perform a simulation (hereafter the MESH run) using a three-dimensional grid-based hydrodynamic code \citep{Mizuta2026}, which builds upon the methods described in \citet{Mizuta2006} and is configured to replicate the physical setup of our fiducial run as closely as possible.
Because such hydrodynamic simulations are computationally expensive, we set up the computational domain in spherical coordinates $(r, \theta, \phi)$ with logarithmic spacing in the radial direction.
 Exploiting the symmetry of the system, we simulate only half of the sphere by restricting the azimuthal domain
to $-\pi/2\le \phi \le \pi/2$ while keeping the full polar range
$0\le\theta\le\pi$. 
The jet axis is placed at $(\theta, \phi)=(\pi/2, 0)$, corresponding to the Cartesian $x$-axis, in order to avoid numerical issues associated with the polar coordinate singularity.
However, for visualization purposes, we transform the coordinates so that the jet is aligned with the $z$-axis, ensuring consistency with the coordinate system adopted in our SPH simulations.
The radial computational domain spans from $r=1.78\times10^{22}\,\mathrm{cm}=5.78\,\mathrm{kpc}$ to $r=4.62\times10^{24}\,\mathrm{cm}=1.50\times10^3\,\mathrm{kpc}$ to align with the SPH setup. 
Reflective boundary conditions are imposed at the inner radial boundary as well as at the angular boundaries in $\theta$ and $\phi$. An outflow (free) boundary condition is applied at $r=r_{\rm max}$. 
The mesh resolution is given by $\Delta r = r\Delta \theta$, with
$\Delta \theta =\Delta \phi =0.5^{\circ}$, ensuring approximately isotropic resolution near the jet axis. 
At the jet launching radius ($r = 10\,\mathrm{kpc}$), this corresponds to a spatial resolution of 0.087 kpc. 
The total number of grid cells is $(N_r, N_{\theta}, N_{\phi}) = (640, 360, 360)$.
We adopt an ideal-gas equation of state with a constant adiabatic index $\gamma = 5/3$.
Although the original code was developed for relativistic jet simulations, it is employed here in the Newtonian regime.

\subsection{Measurements} 
\label{subsec:measurements}

When benchmarking our simulations against analytic models, it is essential to adopt clear, internally consistent measurement definitions. Because we have complete knowledge of the three-dimensional gas distribution and its time evolution, we can construct precise, physically motivated diagnostics.  Throughout this study, we employ uniform criteria to identify different structures (jet lobes, bow shocks) of the jet.  Using these rigorous, simulation-specific definitions ensures that comparisons with theoretical scalings are both meaningful and reproducible.

\subsubsection{Jet lobe}
\label{subsubsec:measure-jet-lobe}

In our simulations, we identify jet-lobe as gas that is simultaneously (i) less dense than the initial ambient medium and (ii) hotter than that medium.  In practice, the lobe is most cleanly isolated in entropy space, where it appears as a high-entropy tail well separated from the shocked ambient gas. 
For a given gas particle, we define the entropy ($K$) with internal energy per unit mass ($u$),  mass density ($\rho$), and the adiabatic index $\gamma=5/3$ \citep{springel_cosmological_2005}
\begin{equation}
    K = \frac{P_{\mathrm{th}}}{\rho^\gamma} =  (\gamma-1) \frac{u}{\rho^{\gamma-1}}
\end{equation}
where $P_{\mathrm{th}} = (\gamma-1)\rho u$ is the thermal pressure. 
We then adopt an entropy threshold
\begin{equation}
    \log_{10} K_\mathrm{t} = \log_{10} K_{0} + f_\mathrm{t} (\log_{10} K_\mathrm{max} - \log_{10} K_{0}),
\end{equation}
where $K_0 = 4.24 \times 10^{32}\,\mathrm{cm^4\cdot g^{-2/3}\cdot s^{-2}}$ is the initial ambient entropy, $K_\mathrm{max}$ is the maximum entropy in the domain at the time of measurement, and $f_\mathrm{t}=0.5$ sets the threshold halfway (in log-space) between the ambient value and the absolute maximum.


\subsubsection{Bow shock}
\label{subsubsec:measure-bow-shock}

As the jet propagates into the surrounding medium, it drives a strong shock into the initially static ambient gas.  Gas swept up by the shock is first accelerated to supersonic speeds, then decelerates to transonic and finally subsonic flow as it expands away from the jet head.  The swept-up layer forms a bow shock that envelops the jet lobe and its surrounding gas.

In our diagnostics, this bow shock is most clearly visible in Mach-number maps, where it appears as a sharp jump in $\mathcal{M}$ from the essentially stationary ambient gas to the transonic medium.  We delineate the shock front by selecting all gas with
$\mathcal M > 0.0316 \quad (|v| > 10^{-1.5}\,c_{\mathrm s,0})$.  

For those components, their lengths are defined as half the distance between the two most distant gas particles in opposite directions along the jet axis that satisfy the entropy criterion. The radius measurement is more sensitive to turbulence along the volume boundary. To obtain a stable estimate, we compute the mean cylindrical radius of the outermost 10 per cent of particles, measured relative to the jet axis.  This procedure filters out small-scale corrugations and yields a robust measure of the radial extent. 
For our MESH run, we used a similar criteria; however, the size measurement is derived from pixel masks taken from a slice in the $x$-$z$ plane. 

\section{Results}
\label{sec:results}

We begin by presenting the results of our fiducial simulation run (FID) in Section~\ref{subsec:fiducial-jets}, followed by a systematic comparison with the suite of simulation variants in Section~\ref{subsec:compare-diff-jets}.

\subsection{Fiducial jets}
\label{subsec:fiducial-jets}

This section analyzes the fiducial jet simulation, denoted as FID in Table~\ref{table:list-of-simulation}. The set of adopted parameters yields characteristic length scales of $L_1 = 32.9\,\mathrm{kpc}$ and $L_2 = 915\,\mathrm{kpc}$, which define the initial and asymptotic regimes of the jet evolution, respectively. These serve as key reference points for interpreting the dynamics, energetics, and morphology of the jet and its bow shock throughout the simulation.

\subsubsection{Time evolution of the jet morphology}

\begin{figure*}
\centering
\includegraphics[width=\textwidth]{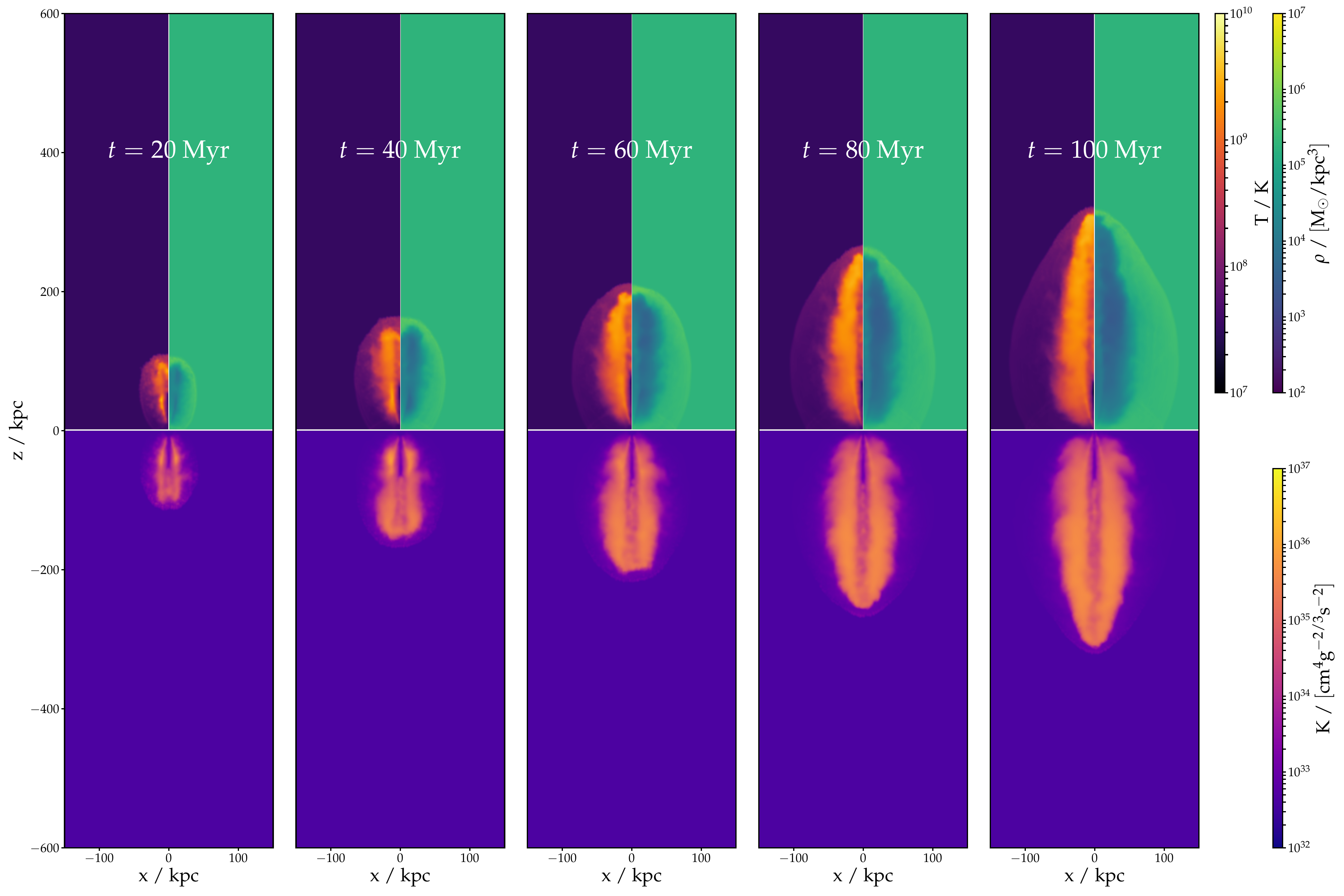}
\caption{\small{
Time evolution of the gas temperature and density for the Fiducial jet run, shown in slices with a thickness of 10 kpc. The simulation employs a time-dependent artificial viscosity scheme and a half-opening angle of $\theta_\mathrm{j} = 10^\circ$ using the Reservoir Grid (RG) launching model. For further parameters, refer to Table~\ref{table:list-of-simulation}.
}}
\label{fig:fid_temp_dens_evo}
\end{figure*}

Figure~\ref{fig:fid_temp_dens_evo} presents the time evolution of the gas temperature, density, and entropy slices for the fiducial (FID) run. 
At $t=20\Myr$, the jet reaches a height of approximately 100 kpc, and the morphological features outlined in Figure~\ref{fig:jet-schematic} become evident.  
The central jet lobe exhibits a low-density, low-temperature, and low-entropy zone near the launch site, surrounded by a higher-temperature envelope. 
This central region contains newly launched particles with high velocity that have not yet significantly interacted with the ambient medium.

At larger vertical distances, the ejected particles begin to interact with the surrounding gas, converting kinetic energy into thermal energy. 
The resulting jet head, enclosed by the bow shock, shows elevated temperature and entropy relative to the unshocked ambient medium.

By $t = 40\Myr$, the overall morphology remains broadly similar to that at $t = 20\Myr$. However, by $t = 60\Myr$, the tip of the jet lobe appears sharper on both the density and temperature maps.
At $t = 80\Myr$, this sharpening extends to the bow shock front, leading to deviations from the expected self-similar evolution. 
At the end of the simulation at $t \sim 100\Myr$, the jet extends to a length of 300 kpc, and the shock morphology of the lobe and bow clearly departs from the self-similar analytic profile.

A notable feature appears near the jet origin: while the central region initially retains a thermal state nearly identical to that of the undisturbed ambient gas, it gradually exhibits signs of jet-ambient interaction. 
By $t = 100\Myr$, two distinct high-entropy streams emerge near the center. This feature reflects the reservoir setup in the RG launching scheme: particles are arranged in layered grids and launched in randomized order per layer. 
Inner-layer particles do not start to launch until the outer layers are depleted. 
Since the reservoir extends to 10 kpc in height, the ambient particles within this region initially remain unperturbed (given a typical smoothing length $h=3.6\,\mathrm{kpc}$ of the ambient gas particles), but are gradually influenced as the inner particles begin to launch.

In terms of quantitative morphology, the jet lobe length and bow shock shape in the FID run generally follow the analytic scaling laws proposed by \citet{Kaiser2007}, particularly during the intermediate stages ($t = 20$-$60\Myr$), where the jet propagation approximately follows the $L_J \propto t^{3/5}$ relation. 
However, at later times ($t \gtrsim 80\Myr$), deviations become apparent as the bow shock narrows and the bow shock front steepens beyond the expected aspect ratio. These changes likely reflect non-linear effects such as the build-up of pressure gradients near the jet head, interaction-induced asymmetries, and resolution-dependent turbulence dissipation. 
The progressive deviation from the self-similar morphology suggests that jet propagation in a stratified medium can enter a kinetically dominated regime once mixing and backflows are suppressed.

The jet exhibits overall symmetry along both the $x$- and $z$-axes, indicating that the RG scheme provides a physically reasonable and stable jet launching strategy.

\subsubsection{Gas characteristic in the simulated jets}
\label{subsubsec:gas-characteristic}

\begin{figure*}
\centering
\includegraphics[width=\textwidth]{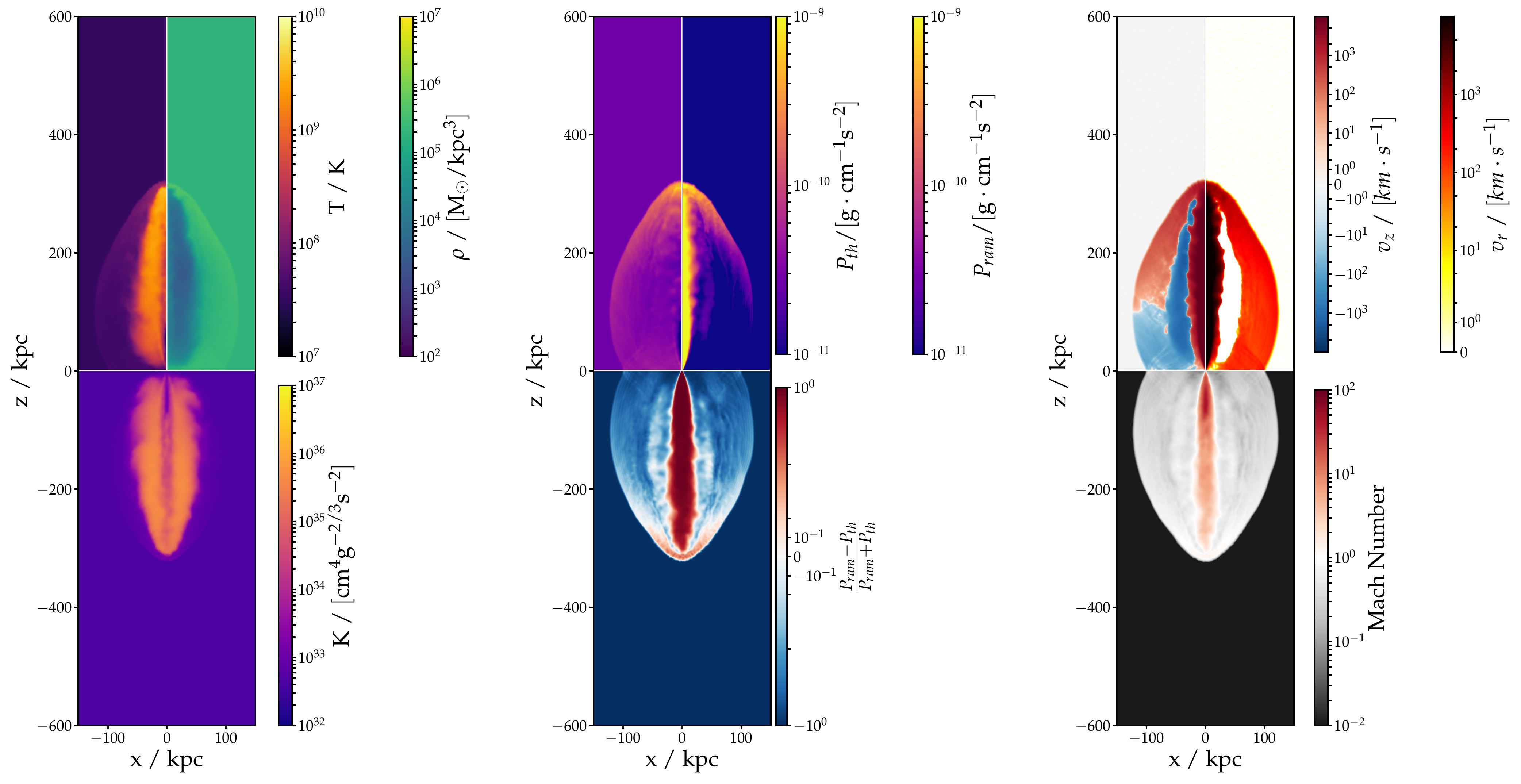}
\caption{\small{
Gas properties at $t = 100\,\mathrm{Myr}$ for the fiducial run. 
Left panel: temperature, density, and entropy slices reveal thermodynamic signatures of the jet lobe and shocked ambient gas. 
Middle panel: thermal pressure, ram pressure, and pressure contrast highlight the jet spine and the turbulent interface with the ambient medium. 
Right panel: vertical velocity ($v_z$), radial velocity ($v_r$), and Mach number diagnose flow kinematics, including supersonic jet material, subsonic lobe backflows, and stationary ambient gas. The movie of these diagrams are available at \url{https://www.youtube.com/watch?v=GfCrq3X46Ag}.
}}
\label{fig:fid_allmapplot}
\end{figure*}

Figure~\ref{fig:fid_allmapplot} presents a series of slice plots showcasing key gas properties at the end of the fiducial run ($t = 100\Myr$). These maps allow us to identify the distinct physical components of the jet structure;  namely, the jet material, the lobe, and the bow shock, through their thermodynamic and kinematic signatures.

In the left panel, we show the temperature, density, and entropy distributions. 
The jet lobe appears as a hot, low-density cavity with markedly elevated entropy compared to that of the ambient medium. 
This region is the clearest thermodynamic indicator of the jet influence. 
Notably, there is no sharp boundary between the lobe and the jet material, which together form a continuous structure of entrained and injected gas. 
Outside the lobe, a broader region of shocked ambient gas is barely visible, enveloped by the bow shock front. 
The surrounding region exhibits an enhanced density and temperature as a result of the shock gas being compressed and heated by the expanding jet.

The middle panel of Figure~\ref{fig:fid_allmapplot} displays thermal pressure, ram pressure, and the contrast between the two. 
These quantities are effective in distinguishing between different jet-driven regimes. The jet material near the axis is characterized by a high ram pressure and moderate thermal pressure, producing a pronounced pressure contrast. 
This region also exhibits a recollimation pattern, the so-called jet spine, where the ram pressure peaks as the flow tightens along the axis. The jet maintains a nearly uniform, high-velocity stream along the axis until it decelerates abruptly at the jet head. 
In contrast, the shocked ambient gas surrounding the spine shows ram and thermal pressures of comparable magnitude. 
In this region, the gas exhibits a bulk velocity that approximates the local sound speed as it moves away from the jet axis, creating a distinct boundary that is differentiated from the surrounding, non-turbulent ambient gas. 
Although the jet lobe can be traced in this panel, its boundaries are less clearly defined than those in the thermodynamic maps. 
It appears as an area within the shock front where the thermal pressure and ram pressure are nearly equivalent.

The right panel of Figure \ref{fig:fid_allmapplot} presents the vertical velocity ($v_z$), the radial velocity ($v_r$) and the Mach number. These kinematic diagnostics effectively separate the different dynamical components of the jet system. 
The ambient medium exhibits negligible velocity, and thus a near-zero Mach number. In contrast, the gas within the shocked boundary moves at approximately sonic speeds. 
The jet material itself is clearly identifiable as a supersonic stream along the axis, registering the highest values of $v_z$ and $v_r$. 
Meanwhile, the jet lobe is particularly well distinguished in the $v_z$ and $v_r$ maps by its characteristic backflow --- regions of negative velocity --- indicating material that is being recycled inward from the edges of the lobe. 
These backflows contribute significantly to the lobe’s pressure support and the redistribution of energy and metals. 
The Mach number map confirms that the lobe material is near-sonic speed, distinguishing it from the supersonic spine and the sub-sonic ambient gas.
The jet material that follows the jet axis is characterized by the highest values of $v_z$ and $v_r$ in the velocity maps, making it the only supersonic component observable in the Mach number map.

\subsubsection{Phase diagrams} \label{subsubsec:phase-diagram}

\begin{figure*}
\centering
\includegraphics[width=\textwidth]{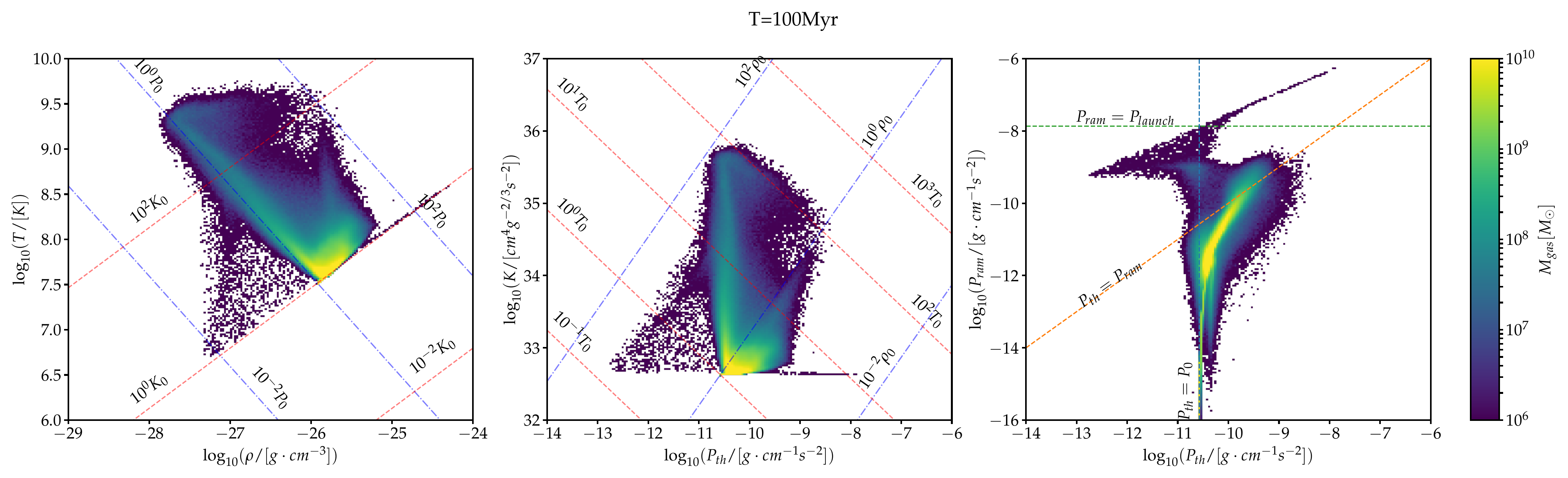}
\caption{\small{
Phase diagrams for the fiducial run at $t \sim 100\,\Myr$: density--temperature ($\rho - T$) plane (left), thermal pressure--entropy ($\pth - K$) plane (middle), and thermal--ram pressure ($\pth - \pram$) plane (right). 
In the left panel, isobaric (blue dot-dashed lines) and adiabatic (or isentropic, red dashed lines) curves are shown. 
The middle panel shows isodensity (blue dot-dashed) and isothermal (red dashed) lines. 
In the right panel, the red dashed line marks $\pth = \pram$, the vertical blue dot-dashed line indicates the initial ambient pressure, and the green long-dashed line shows the ram pressure of jet material at launch.
The movie of this diagram is available at \url{https://www.youtube.com/watch?v=YwKruJ1pR3A}. 
}}
\label{fig:fid_phase_digram_all}
\end{figure*}

Figure~\ref{fig:fid_phase_digram_all} presents a set of phase diagrams at $t \sim 100\,\Myr$ for the fiducial run, providing thermodynamic and kinematic diagnostics of the gas impacted by the jet. 
We show the gas distribution in the density-temperature ($\rho - T$) plane (left), thermal pressure--entropy ($\pth - K$) plane (middle), and thermal--ram pressure ($\pth - \pram$) plane (right). The color scale indicates the mass density of the gas in each phase space bin.

The $\rho - T$ diagram and $\pth - K$ diagram are mathematically related via affine transformations, but we display them separately to highlight different aspects of gas thermodynamics. 
In both panels, one can identify a strong population aligned with the ambient pressure level ($P \approx P_0$), extending over a wide entropy range ($K_0 < K < 10^3 K_0$). 
This distribution reflects the thermodynamic evolution of the shocked ambient gas and the lobe material. 
A secondary feature appears along the adiabatic (isentropic) relation, with modest enhancement in entropy beyond $K_0$, corresponding to gas undergoing expansion with limited shock heating.

A distinct subpopulation at constant density and elevated temperatures ($T_0 < T < 10^2\,T_0$) is also visible, associated with jet head where the shocked gas exchanges internal energy with the ambient gas. 
In contrast, a diffuse `delta-shaped' region in the lower pressure, high entropy quadrant ($P < P_0$, $K > K_0$) represents the newly launched jet gas that endures a nearly free expansion in the jet spine. 
Another diffuse component spans the regime $P_0 < P < 10^2\,P_0$, indicating the presence of partially mixed material at intermediate pressures.

The right panel shows the phase diagram in the thermal-ram pressure plane, which offers a diagnostic of jet-ambient interactions.
In the phase diagram, two branches emerge. 
One passes through the intersection of $\pth = P_0$ and $\pram = P_{\mathrm{launch}}$, whose slope matches the adiabatic process assuming a constant speed. 
This branch corresponds to recently launched, supersonic jet material, which moves without much deacceleration and follows adiabatic expansion. 
The second branch follows $\pth \approx P_0$ at small $\pram$ values, but turns to align with $\pth \sim \pram$ when $\pram$ goes higher.
This population represents the ambient gas: the low $\pram$ part following $\pth \approx P_0$ is the shock heated ambient gas, and the $\pth \sim \pram$ trend marks the ambient gas in the jet lobe that interacts with the jet particles. 
The region where two branches meets near $(\pth, \pram) \sim (P_0, 10^{-1} P_{\mathrm{launch}})$ marks the transition regime where the shocked jet material mixes with the ambient medium. 
This structure provides insight into the evolution of pressure support and turbulence inside the bow shock.



\subsubsection{Jet lobe decomposition}
\label{subsubsec:jet-lobe-decomposition}

\begin{figure*}
\centering
\includegraphics[width=\textwidth]{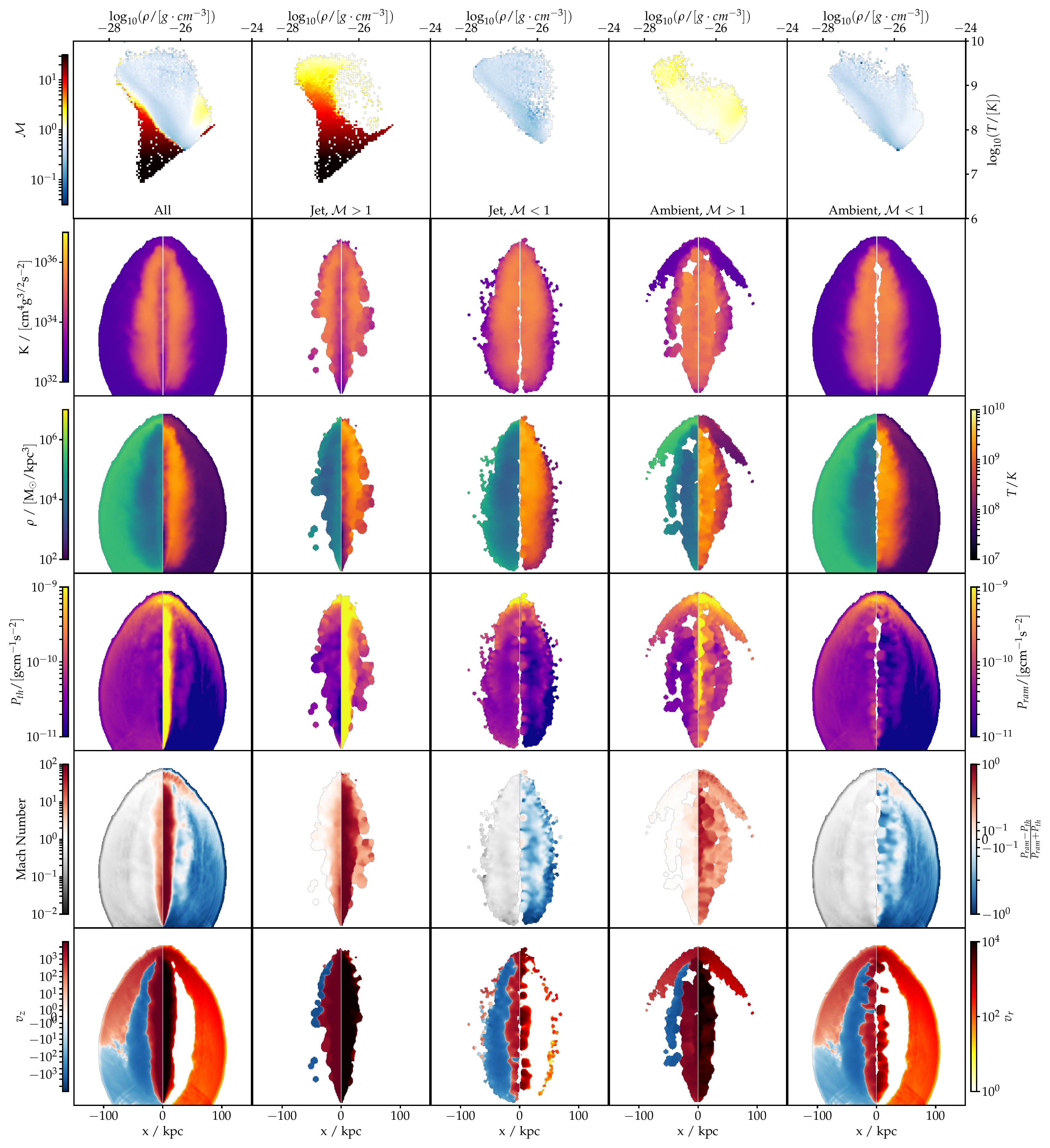}
\caption{\small{Gas properties of the jet lobe decomposed into the jet gas and the shocked ambient gas with high Mach number ($\mathcal{M} > 1$) and low Mach number ($\mathcal{M} < 1$).}
From left to right, the figure shows the slices of all the shocked gas, supersonic jet gas, subsonic jet gas, supersonic ambient gas, and subsonic ambient gas.
From top to bottom, the panels are the $\rho$--$T$ phase diagram color-coded with Mach number, entropy slices, density--temperature slices, thermal--ram pressure slices, Mach number--pressure contrast slices and $v_z$--$v_r$ slices, respectively. 
}
\label{fig:fid_decomposition_allmapplot}
\end{figure*}

\begin{figure*}
\centering
\includegraphics[width=\textwidth]{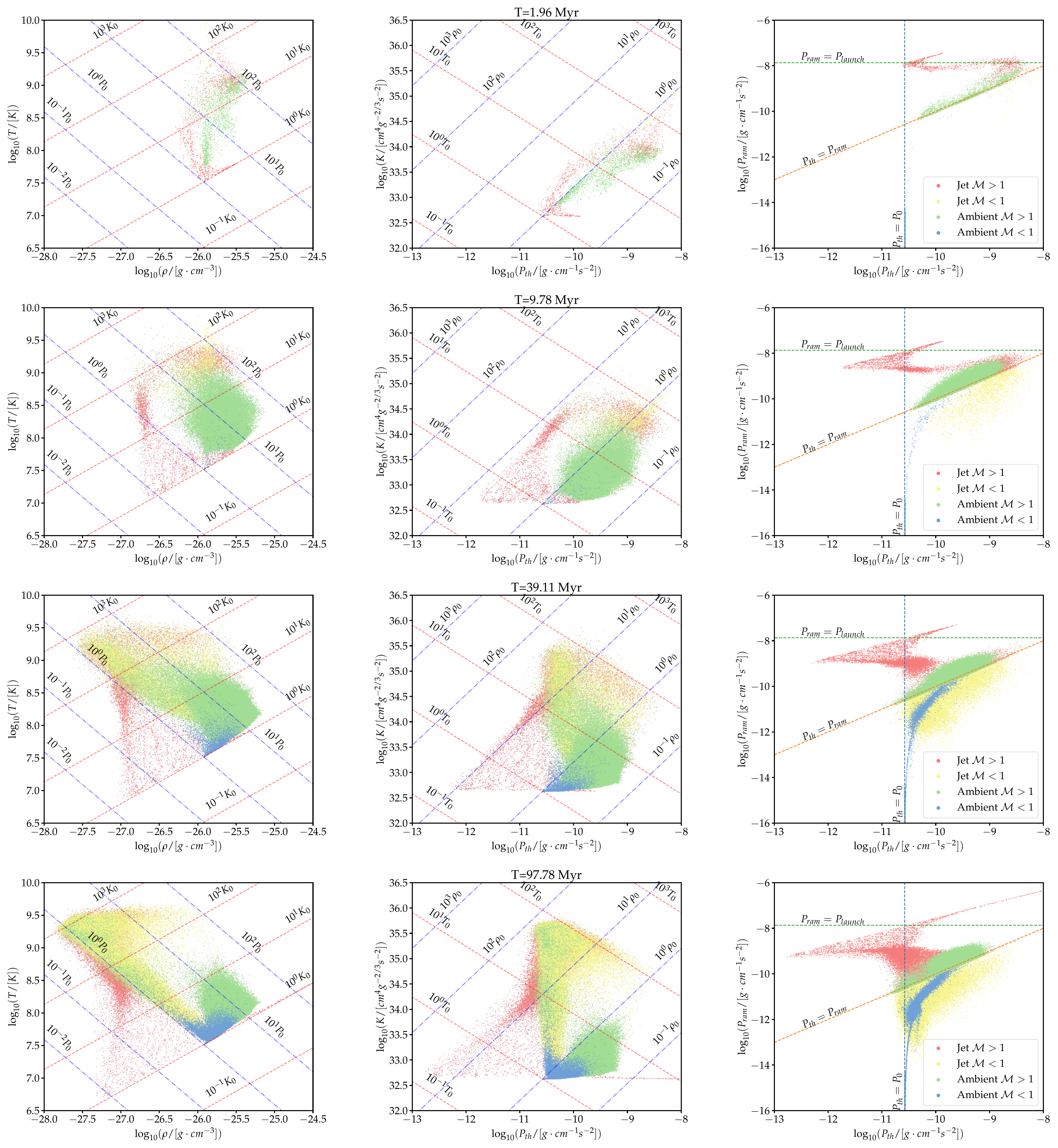}
\caption{Gas phase diagram for components defined in Fig. \ref{fig:fid_decomposition_allmapplot}.
The components are color-coded with red (jet gas with $\mathcal{M}>1$), yellow (jet gas with $\mathcal{M}<1$)), green (ambient gas with $\mathcal{M}>1$), and blue (ambient gas with $\mathcal{M}<1$). 
From left to right, the panels are the phase diagram on the density-temperature ($\rho - T$) plane, thermal pressure--entropy ($\pth - K$) plane, and thermal-ram pressure ($\pth - \pram$) plane, respectively. 
From top to bottom are the snapshots at $t=1.96, 4.89, 9.78, 39.11, 97.78\Myr$.
The movie of this figure is available at \url{https://youtu.be/n933dMN-YcE}.
}
\label{fig:fid_decomposition_scatter}
\end{figure*}

The findings in sections \ref{subsubsec:gas-characteristic} and \ref{subsubsec:phase-diagram} motivate us to decompose the jet into the approach detailed in Section \ref{subsec:measurements}. 
We apply a filter to the snapshot using the criteria for the shocked gas ($\mathcal{M}>0.0316$) to remove ambient unshocked gas, then further classify the gas based on two criteria: gas origin (from jet ejection or ambient) and jet speed (subsonic or supersonic, determined by Mach number).
Figures \ref{fig:fid_decomposition_allmapplot} and \ref{fig:fid_decomposition_scatter} are the slice plots and the phase diagrams for jet decomposition, respectively. 

The jet particles with $\mathcal{M}>1$ (second column in Figure \ref{fig:fid_decomposition_allmapplot}; red points in Figure \ref{fig:fid_decomposition_scatter}) correspond primarily to the particles ejected in the jet spine. 
These particles are dispersed across the $\rho$-$T$ plane; near the jet axis, a substructure with low entropy, low temperature, high ram pressure and high Mach number is observed. 
These particles have minimal interaction with the ambient gas, resulting in negligible kinetic energy loss. 
At larger radii, jet particles exhibit higher entropy and lower ram pressure, indicating interaction with the surroundings and a reduction in speed. 
A gradient in the Mach number is observed along the temperature axis on the phase diagram, as anticipated: initially, jet particles possess high velocity (high Mach number) and low temperature, then decelerate and heat up upon interaction.
From the $\pth$-$K$ phase diagram, it is clear that this population is situated at the isobaric component and the `delta region.' The $\pth$-$\pram$ plot shows the inclusion of the adiabatic component, as depicted in Figure \ref{fig:fid_phase_digram_all}.

Jet particles with $\mathcal{M}<1$ (third column in Figure \ref{fig:fid_decomposition_allmapplot}; yellow points in Figure \ref{fig:fid_decomposition_scatter}) predominantly coincide with the structure of the jet lobe. 
This group is characterized by high temperature, low density, and high entropy, suggesting the low-density volume evacuated by the jet particles. 
Notably, the vertical velocity of these particles presents a backflow with speeds of several thousand km/s. 
The particles also have a notable negative radial velocity. 
These negative velocities suggest that low-Mach-number jet particles have reached the tip of the jet and are being pushed back because of the gas pressure. 
In the phase diagram, these particles sit along the same equal-pressure line as the $\mathcal{M}>1$ component. 
However, there is a marked difference in the Mach number, particularly at the low-entropy end: a location on the $\rho$-$T$ diagram can simultaneously exhibit $\mathcal{M}>10$ and $\mathcal{M}\sim0.1$.

The component of ambient gas with $\mathcal{M}>1$ (fourth column of Figure \ref{fig:fid_decomposition_allmapplot}; green points in Figure \ref{fig:fid_decomposition_scatter}) interacts with high-speed jet particles or participates in jet propagation. 
These particles exhibit similar thermal conditions as the jet particles but have lower Mach numbers. 
Compared with their spatial distribution to the jet core, these particles form the shock `head,' which is characterized by low entropy, high temperature, and high density, in addition to the jet lobe. 
In the phase diagram, the `jet lobe' aligns with $P=P_0$, similar to the jet gas with $\mathrm{M}<1$, while the `jet head' is closer to the original ambient gas $(\rho_0, T_0)$ and is located in the low-entropy region of the $\pth$-$K$ diagram.

The most substantial content in the bow shock is the ambient gas with $\mathcal{M}<1$. In Section \ref{subsubsec:phase-diagram}, we identified three primary components, with the most prominent having an entropy just above $K_0$.

Combining the aforementioned analysis, we can trace the evolution of jet particles in phase space. Initiated at $(\rho_0, T_0)$, the jet particle has a high velocity, as evidenced by an elevated Mach number. 
Initially, dependent on the previous cavity scale, the particle experiences near-free expansion until reaching an appropriate smoothing length, putting it in the `delta region'. 
During this process, heat convection with surrounding particles and interaction-driven deceleration occur, resulting in a Mach number gradient (see the top panels of Figure \ref{fig:fid_decomposition_scatter}). 
The particle then gradually approaches the isopressure line $P=P_0$ due to the interaction with jet lobe material. 
As the high-speed particle collides with the hot jet head, its speed drops. Meanwhile, density and temperature drastically increase as a result of shock compression, converting kinetic energy into thermal energy. 
This process shifts the particle toward the upper right phase diagram into another diffuse region.
In real space, particles rebound from the tip, residing within the jet lobe. 
Here, the particle decelerates to subsonic speeds, losing internal energy to the ambient gas. 
The final state is achieved when the local pressure equilibrium is met, aligned with the $P=P_0$ branch and $\mathcal{M}<1$.

In the phase diagram of ambient gas, the primary shaping factors are shock compression and jet particle interactions. 
Unlike the jet particle, the shock front moves through the ambient gas, causing an increase in density, temperature, and entropy, resulting in an iso-entropy component at $K$ slightly above $K_0$. 
The ambient gas's iso-density component arises from heat exchange at the jet head, where super-heated jet particles cool after shock compression. 
Ambient gas particles enclosed by the shock front reach pressure equilibrium with their surroundings, forming a component along the iso-pressure line.


\subsubsection{Jet size evolution}

\begin{figure}
\centering
\includegraphics[width=\columnwidth]{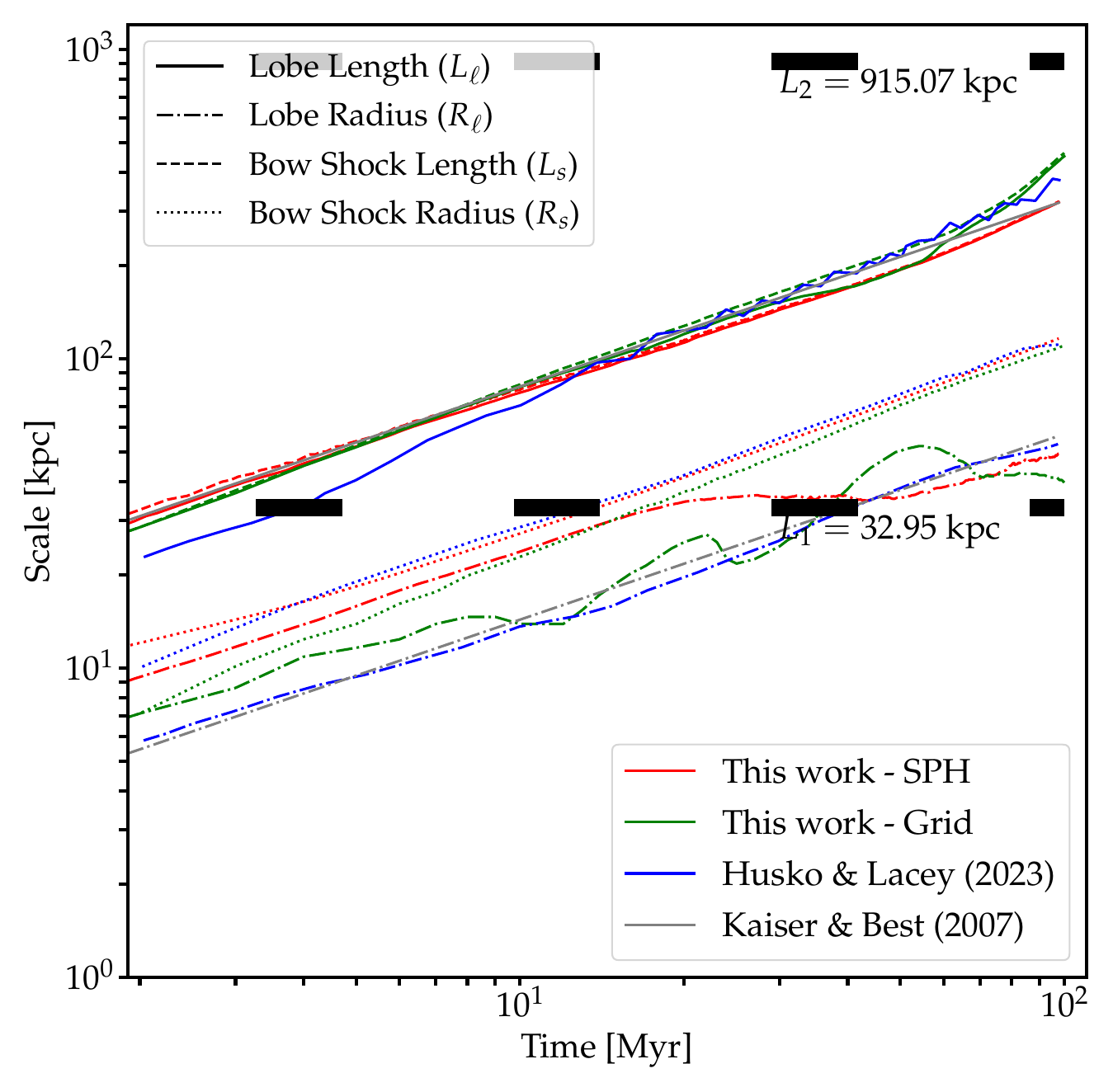}
\caption{\small{
Time evolution of jet-driven structure sizes in this work. 
The red line and green lines represent the size measurement of FID and MESH run in this work, respectively. Blue lines denote the simulation results of \Husko{}, and gray lines show the analytic prediction from \citet{Kaiser2007}. 
Solid, dot-dashed, dashed, and dotted lines represent the lobe length ($L_\ell$), lobe radius ($R_\ell$), bow shock length ($L_s$), and bow shock radius ($R_s$), respectively. 
The values of $L_1$ and $L_2$ computed using Eqs.~(\ref{eq:L1}) and (\ref{eq:L2}) are marked with black wavy lines.
}}
\label{fig:fid_lobesize_evo}
\end{figure}

Figure~\ref{fig:fid_lobesize_evo} presents the time evolution of the characteristic spatial scales in our fiducial jet simulation, including the lengths and radii of the lobe and bow shocks. These are compared against the analytic self-similar model of \citet{Kaiser2007} and the simulation results of \Husko{}. 
We also included the measurement of MESH run as green lines in the figure, which is discussed later in Section \ref{subsec:grid_sim}.

We define the extent of the lobe using an entropy-based criterion (\S\ref{subsubsec:measure-jet-lobe}), which reliably tracks the thermalized, low-density cavity produced by jet activity. 
Overall, our measured lobe length (solid red line) follows the theoretical scaling $L_J \propto t^{3/5}$ reasonably well across most of the simulation, confirming that the the expansion of bow shock broadly adhere to self-similar expansion. 
However, at late times ($t \gtrsim 70$ Myr), a mild deviation appears: the lobe length curve flattens slightly, diverging by $\sim$10\% from the analytic curve. This may reflect residual turbulence, finite-resolution effects, or asymmetry in the lobe structure near the termination phase.

In contrast, the simulation by \Husko{} (blue solid line) displays slightly longer jet lobes at late times, exceeding both our results and the theoretical expectation. This may be due to differing numerical viscosity, launching schemes, or entropy thresholds used to define the jet-ambient interface (cf. \S\ref{subsec:compare-diff-jets}). Notably, our lobe structure remains more compact, suggesting enhanced mixing or greater dissipation of kinetic energy into heat (see Fig.~\ref{fig:fid_energy_evo}).

The evolution of the lobe radius ($R_\ell$, dot-dashed red line) initially lags behind the analytic prediction and the \Husko{} result for the first $\sim$20 Myr. During this early phase, the radial width closely tracks the bow shock radius ($R_s$), hinting at resolution limitations that suppress early lateral expansion. After $t \sim 20$ Myr, however, $R_\ell$ begins to converge toward the expected scaling, albeit consistently remaining slightly narrower than the theoretical or \Husko{} profiles.

As with lobe length, delayed growth in lobe width also suggests a departure from self-similar evolution. 
An increase in the lobe aspect ratio (length-to-radius) was predicted analytically by \citet{Alexander:2002} to arise from gravity-driven perturbations. 
In our view, however, this mechanism cannot explain the analogous behavior observed in our simulations, since gravity is not included. 
The grid-based simulation of \citet{Gaibler:2009} found that the aspect ratio increases for a light jet propagating in a uniform ambient medium. 
A direct comparison between our results and those of \citet{Gaibler:2009} is not straightforward, since our setup involves a heavy jet. 
However, we observe that the light jet exhibits a steadily increasing aspect ratio, whereas the heavier jets show a rapid initial rise followed by much slower growth, consistent with the evolution trend seen in our simulation around $t\simeq 20\Myr$.

The bow shock structure, traced via entropy and pressure discontinuities (\S\ref{subsubsec:measure-bow-shock}), reveals a coherent and egg-shaped expansion of the shock front. The bow shock length ($L_c$, dashed red line) remains modestly ahead of the jet lobe throughout, indicative of a persistent contact discontinuity between the forward shock and the jet head. The radial expansion of the shock front ($R_s$, dotted red line) maintains a nearly constant aspect ratio relative to $L_s$, suggesting that the shock front as a whole evolves self-similarly, even when the internal lobe structure deviates due to mixing, turbulence, or dissipation.

Overall, these results validate the applicability of analytic jet models for describing large-scale bubble growth but also highlight the need for high-resolution simulations and entropy-based diagnostics to capture deviations from self-similarity at late times.

\subsubsection{Energy composition evolution}
\label{subsec:fiducial-energy-evo}

\begin{figure}
\centering
\includegraphics[width=\columnwidth]{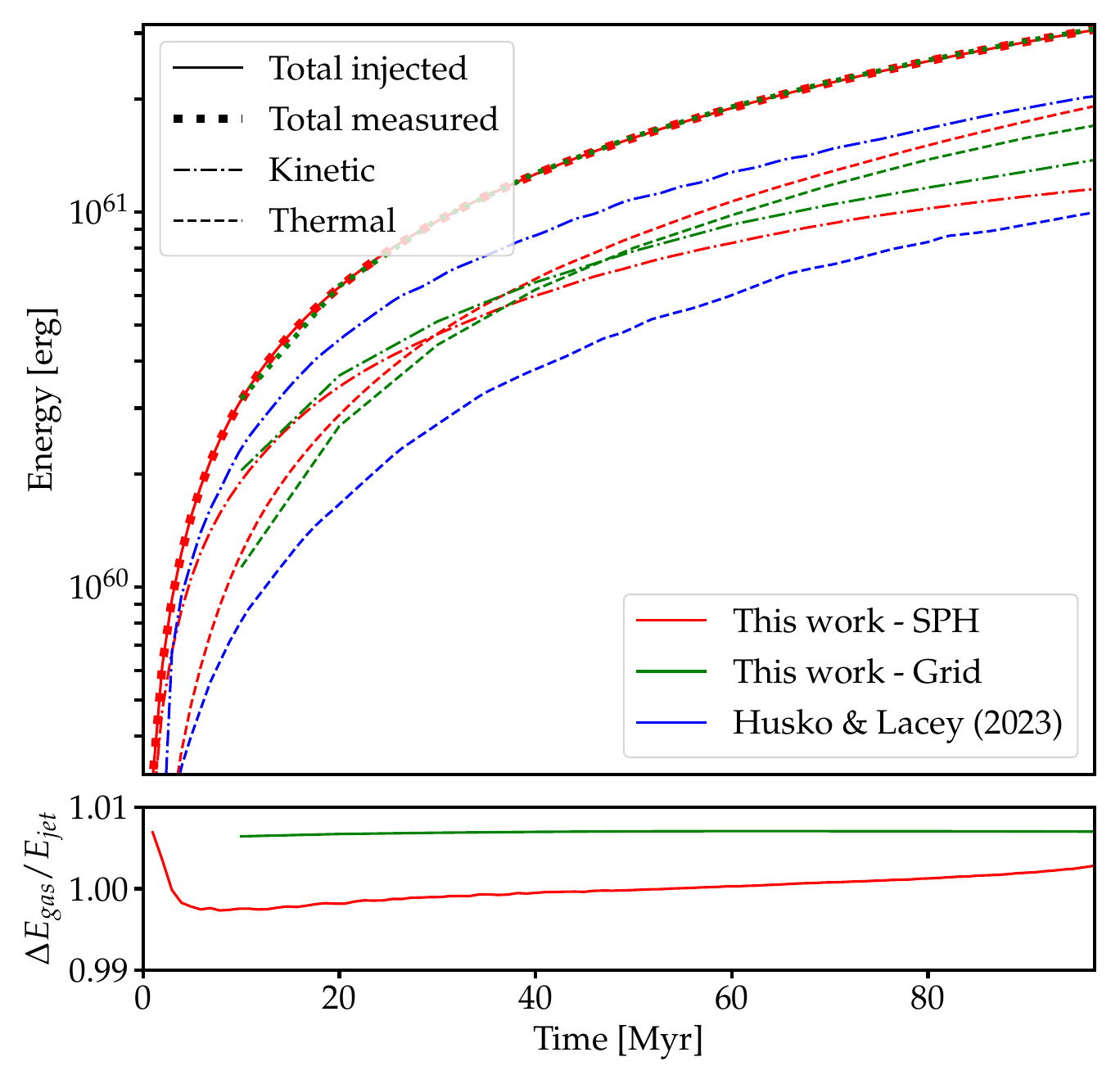}
\caption{\small{
Time evolution of the energy budget in the fiducial SPH jet run. The kinetic (dot-dashed lines) and thermal (dashed lines) components are shown separately, while their sum represents the total measured gas energy (thick dotted lines). 
The total energy injected by the jet is plotted as a solid line for comparison. 
Red curves denote fiducial results from this work; green ones represents MESH run; blue curves show the corresponding evolution from \Husko{} after rescaling. 
Our simulation conserves total energy to within 1\% as shown in the bottom sub-panel, while exhibiting a time-dependent partitioning: kinetic energy dominates early on, but thermal energy grows increasingly prominent after $t \gtrsim 20$ Myr due to shock heating and turbulent dissipation in the shocked ambient gas.
}}

\label{fig:fid_energy_evo}
\end{figure}


Figure~\ref{fig:fid_energy_evo} shows the time evolution of the energy partition in the fiducial run, decomposed into kinetic (dot-dashed line), thermal (dashed line), and their measured sum (thick dotted line). For reference, the total injected energy from the jet source is shown as a solid line.  The red color is used to represent our results, while the blue color indicates the rescaled \footnote{The total energy budget in \Husko{} is $\sim0.05$ dex higher than the prediction of $P_{\mathrm{jet}}=10^{46}\,\mathrm{erg/s}$, so we rescaled the energy budget measured in \Husko{} to eliminate the difference. Rescaling has a minor effect on our conclusions.} results from \Husko{}. Similar to Figure~\ref{fig:fid_lobesize_evo}, we also indicate the energy budget in MESH run, which will be discussed in Section \ref{subsec:grid_sim}.

We find that total energy is conserved to within 1\% throughout the simulation, as confirmed by the near-perfect agreement between the measured (thick dotted) and injected (solid) curves. The bottom panel quantifies this small discrepancy, with $\Delta E_{\rm gas}/E_{\rm jet}$ fluctuating around unity by less than 1\%. This level of consistency confirms that our numerical setup captures shock heating and momentum deposition with high fidelity, with negligible artificial dissipation or numerical loss.

A key distinction from the findings of \Husko{} emerges in the time-dependent breakdown between thermal and kinetic energy. In their simulation (blue lines), the thermal-to-kinetic ratio remains nearly constant throughout the full runtime, indicating a steady energy-conversion efficiency during the jet's evolution. 
Notably, the kinetic energy measured by \Husko{} remains about two times higher than the thermal energy in the whole simulation.
In contrast, our run shows a clear temporal trend: kinetic energy dominates the budget during the initial $\sim$20 Myr, but the thermal component overtakes thereafter and continues to grow more rapidly.

The energy budget was also discussed in previous grid-based jet simulations. 
In \citet{Gaibler:2009}, the fraction of thermal energy was found nearly unchanged throughout the simulation, hinting the self-similarity of jet evolution; however, even for the heaviest jet in their simulation, more than half of the energy ejected is in the form of thermal energy, which is in agreement with our kinetic-thermal energy partition at late stage of the jet evolution. 
\citet{Hardcastle:2013} discussed the energy budget in their Figure 9 and 10. 
They reported that the contribution of thermal energy to the overall energy budget exceeds that of kinetic energy.
The gravitational potential energy remains around $10 \%$ of the total energy, implying the role of gravity stays secondary in this stage of jet evolution.

\subsection{Jet morphology across parameter variations}
\label{subsec:compare-diff-jets}

\begin{figure}
\centering
\includegraphics[width=9cm]{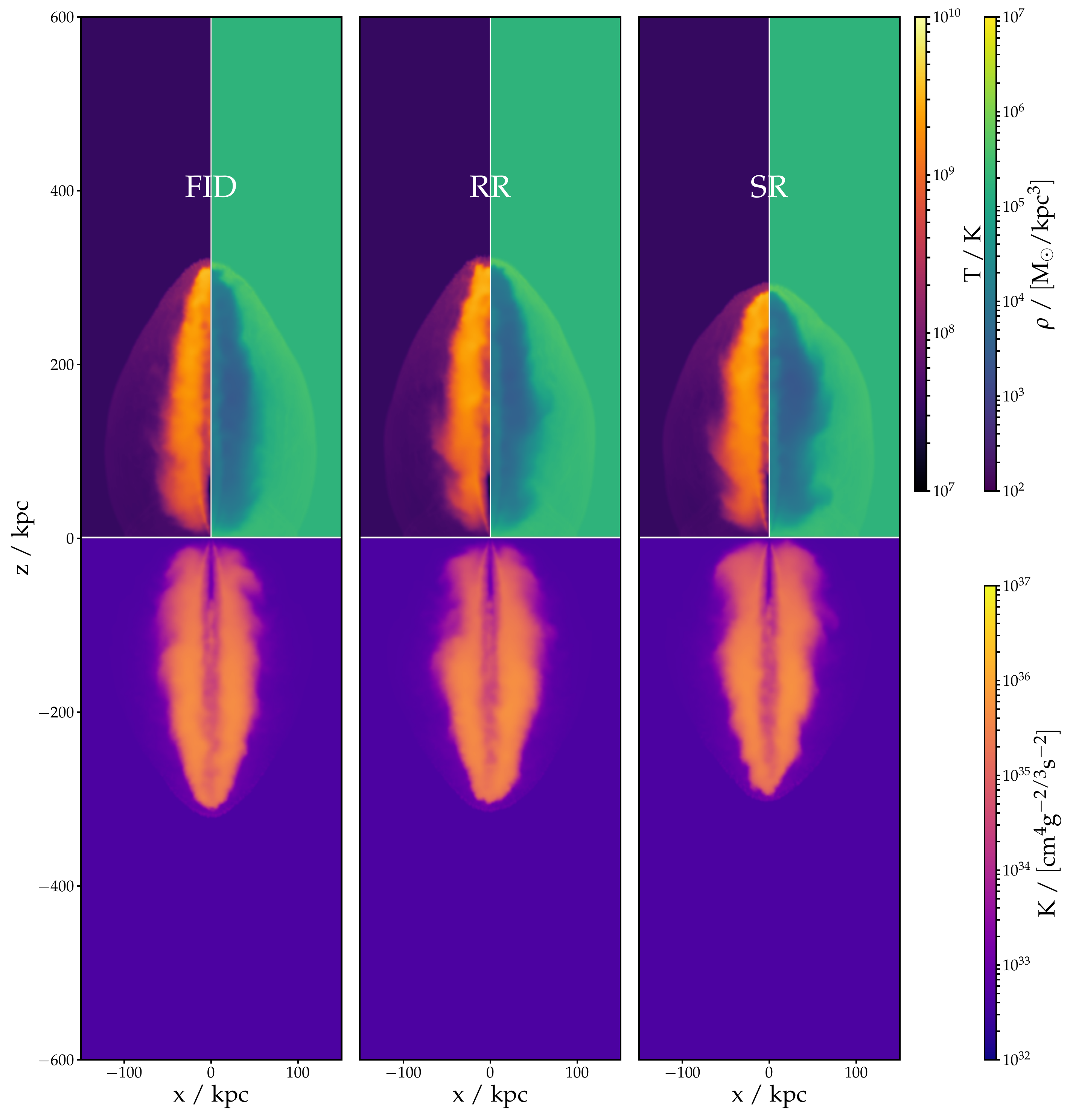}
\caption{\small{
The slice plot of temperature, density, and entropy at the end of the simulation ($t\sim 100\,\Myr$) for different jet particle launching schemes. 
While all three launching schemes show similar jet structure, the SR variant has a shorter lobe. }}
\label{fig:ComparingLaunchingModel}
\end{figure}

\begin{figure}
\centering
\includegraphics[width=\columnwidth]{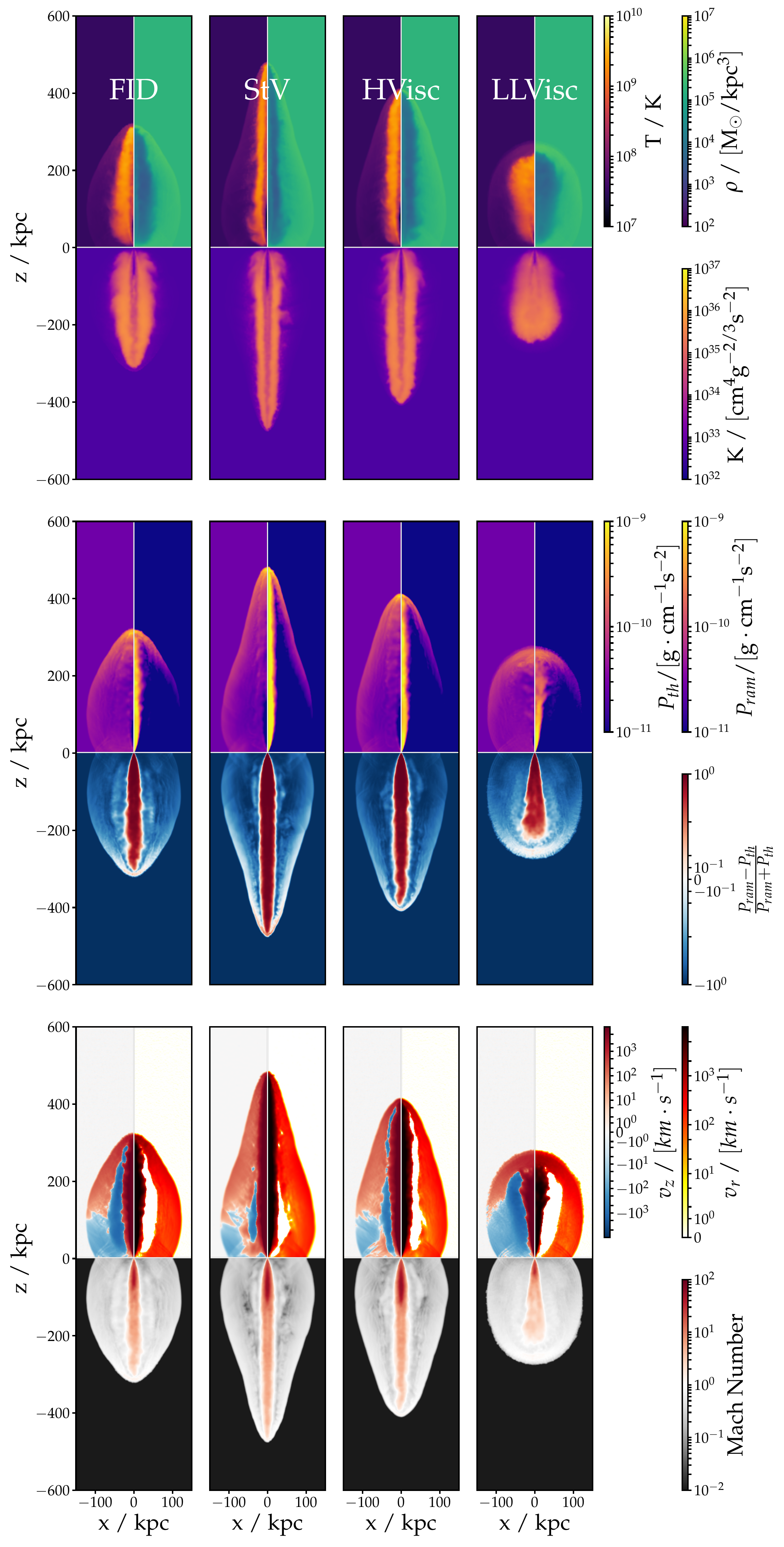}
\caption{\small{
Slice plots at $t \sim 100\Myr$ comparing different artificial viscosity prescriptions: FID (time-dependent viscosity with $\alpha_\mathrm{max}=2$), StV (constant viscosity with $\alpha=2$), HVisc (time-dependent viscosity with $\alpha_\mathrm{max}=5$), and LLVisc (time-dependent viscosity with  $\alpha_\mathrm{max}=0.5$).  
While all runs develop a well-defined jet spine and bow shock, the StV run produces a significantly longer and more collimated lobe. }}
\label{fig:ComparingAV}
\end{figure}

\begin{figure*}
\centering
\includegraphics[width=\textwidth]{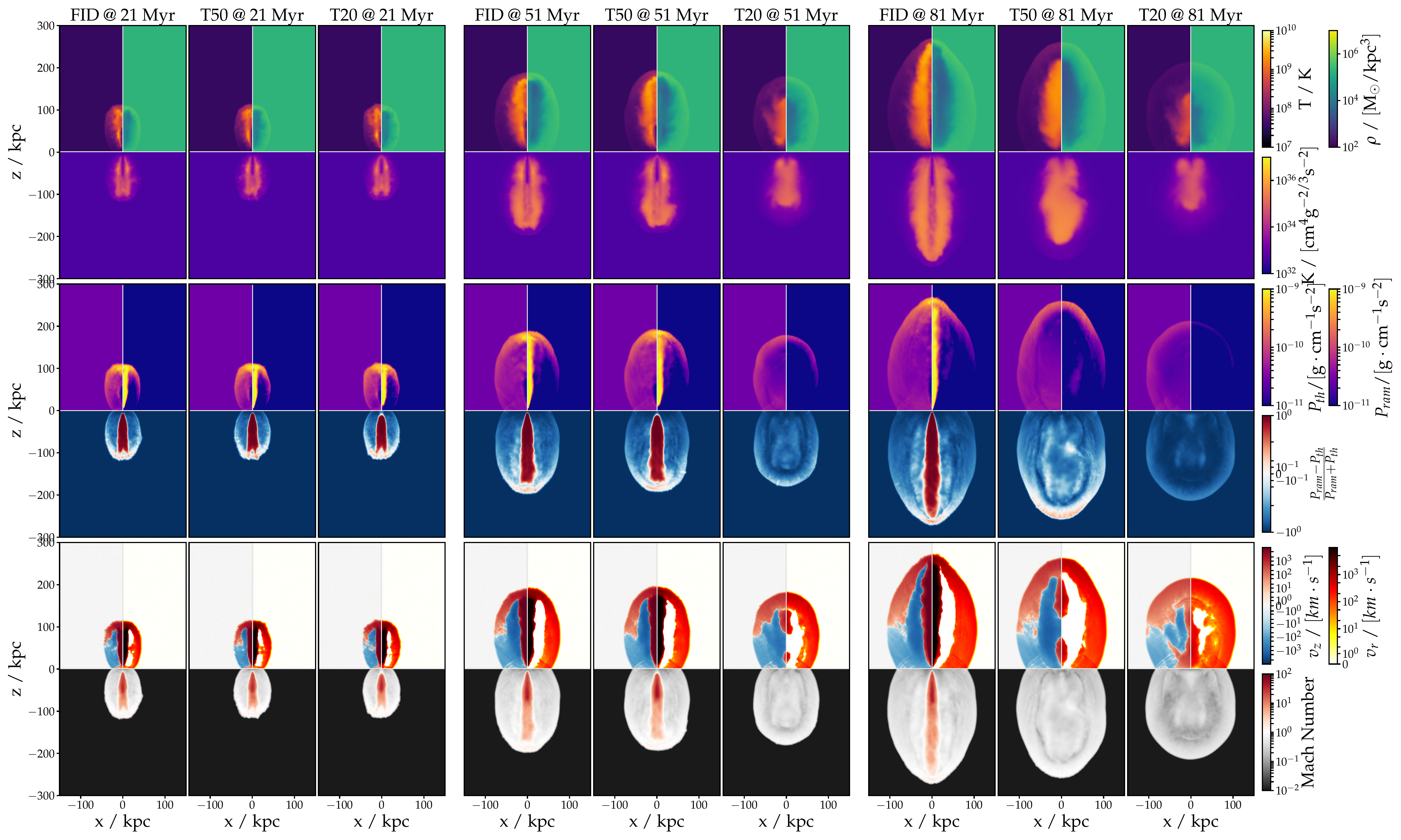}
\caption{\small{
Time evolution of jets with different active lifetimes: FID ($t_{\rm jet}=100$ Myr), T50 ($t_{\rm jet}=50$ Myr), and T20 ($t_{\rm jet}=20$ Myr).  
Each column shows slices of temperature, density, entropy (top row), pressure components (second row), vertical velocity and Mach number (third and fourth rows) at three epochs ($t = 21$, 51, and 81 Myr).  
Despite early jet shutdown in the T20 and T50 runs, key features such as the jet spine, lobe, and bow shock persist for tens of Myr, although their structure gradually dissolves.  
By 30 Myr post-shutdown, the jet spine disappears and lobe growth stalls, while buoyant expansion continues to inflate the bow shock.  
These results suggest that fossil cavities can remain morphologically and thermodynamically distinct long after the energy injection has ceased, providing insight into the nature of observed radio-quiet X-ray bubbles.}}
\label{fig:DTJetVar-TDKPPP}
\end{figure*}

\begin{figure*}
\centering
\includegraphics[width=0.9\linewidth]{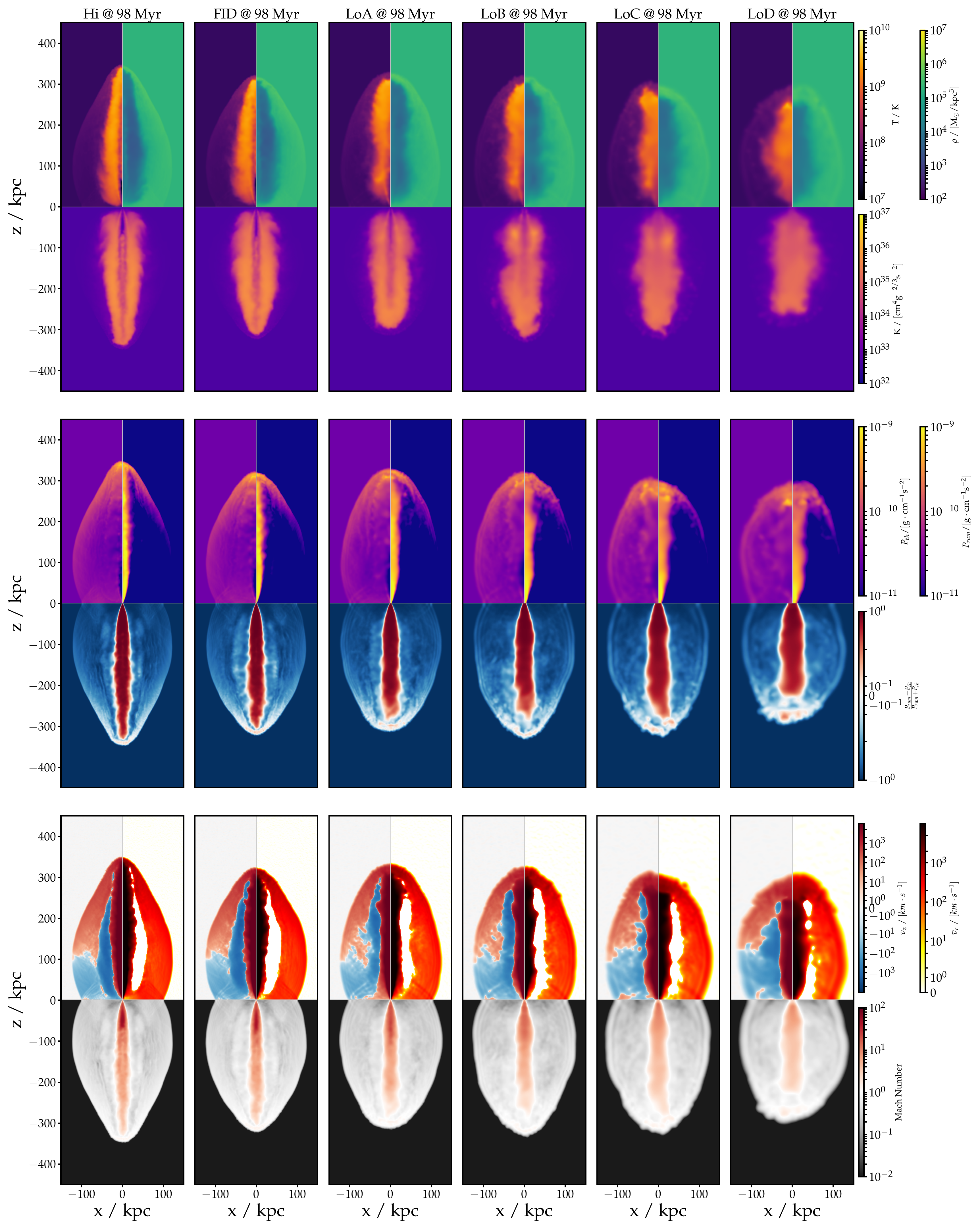}
\caption{\small{
Comparison of jet simulations at different mass resolutions at the end of the simulation ($t = 98\Myr$). 
The number of particles launched into each lobe ranges from $1.21 \times 10^{3}$ (LoD) to $1.51 \times 10^5$ (Hi). These resolutions are representative of typical setups used in cosmological simulations of clusters and large-scale structure. All jet injection parameters are identical across runs. From top to bottom, the panels show slices of gas temperature, density, entropy, pressure components, vertical velocity, and Mach number. While lobe sizes remain similar, structural features such as the jet spine, bow shock sharpness, and internal velocity stratification degrade at lower resolution.}}
\label{fig:Comparison-Resolution}
\end{figure*}

\begin{figure*}
\centering
\includegraphics[width=17cm]{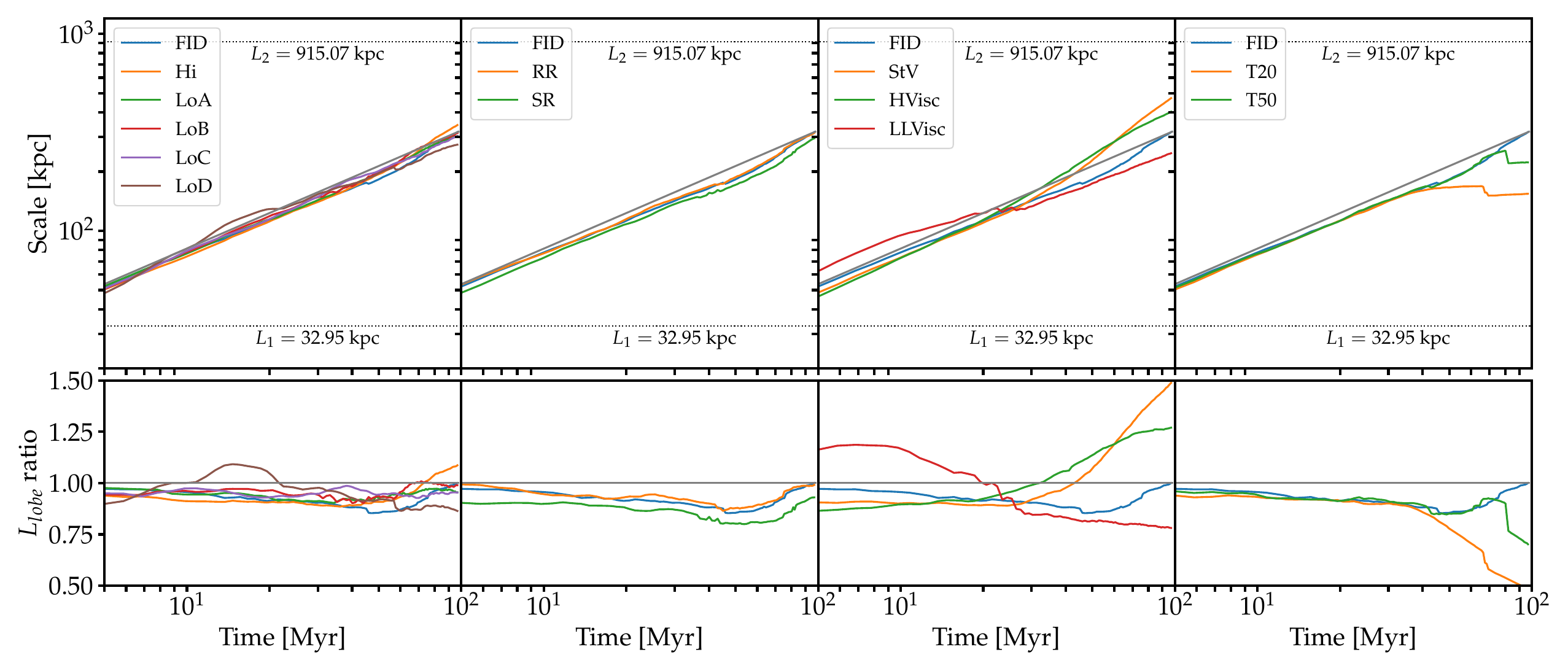}
\caption{\small{
Time evolution of jet lobe length for different simulation variants.  
Top panels show the physical lobe length as a function of time, compared to the analytic scaling $L_J \propto t^{3/5}$ (grey dashed line); bottom panels show the length relative to the fiducial run (FID).  
From left to right, panels correspond to changes in mass resolution (Hi-LoD), launching scheme (RR, SR), artificial viscosity (StV, HVisc, LLVisc), and jet duration (T20, T50).  
Most variants track the fiducial case within $\sim$10\%, but specific configurations—e.g., SR, StV, and truncated jets—show systematic deviations due to altered momentum deposition, mixing efficiency, or post-injection evolution.}}
\label{fig:Comparison-lobe-length}
\end{figure*}

We compare the fiducial simulation (FID) with a series of controlled variants to assess how various numerical and physical parameters influence the morphology, energetics, and evolution of AGN jets. The tested variations include:
(i) the jet particle launching mechanism,
(ii) artificial viscosity scheme,
(iii) jet active duration (duty cycle), and
(iv) mass resolution.
Simulation setups are summarized in Table~\ref{table:list-of-simulation}, and the resulting differences are illustrated through slice plots (Fig. \ref{fig:ComparingLaunchingModel}, \ref{fig:ComparingAV}, \ref{fig:DTJetVar-TDKPPP}, \ref{fig:Comparison-Resolution}) and the time evolution of lobe lengths (Fig.\ref{fig:Comparison-lobe-length}).

\paragraph{Launching mechanism}
Figure~\ref{fig:ComparingLaunchingModel} compares three particle launching schemes: the grid-based reservoir model (FID), randomized launch from the same reservoir (RR), and stochastic spawning at runtime from a fixed radius (SR). While all produce broadly consistent macroscopic morphologies with a central jet spine, expanding lobes, and surrounding gas, subtle yet significant differences arise in their forward propagation.

The FID and RR runs yield similar lobe lengths ($\sim$300 kpc), indicating robustness to random particle ordering as long as the launching geometry is preserved. The SR run, however, lags behind at $\sim$250 kpc. This 15-20\% deficit arises from its injection method: by placing new particles at $r=5$ kpc rather than 10 kpc (as in FID and RR), the SR jet delays the delivery of momentum to the bow shock. Using $v_j\sim10^4$ km/s, the time delay is roughly $0.3$ Myr, which translates to a shorter jet during the linear growth phase that persists into the self-similar regime.

Entropy maps further reveal that the SR lobes are slightly more isotropic and rounded at the tip, consistent with reduced axial thrust and broader angular energy dispersion. These differences highlight that even when the total energy and momentum is conserved, the specific details of how the particles are initialized can affect the collimation, directionality, and eventual observables, particularly in subgrid models of feedback.

\paragraph{Artificial viscosity and jet coherence}
The contrast among FID, StV, HVisc, and LLVisc, as shown in Fig.~\ref{fig:ComparingAV}, demonstrates the critical influence of numerical viscosity. 
FID uses the time-dependent Cullen--Dehnen switch to suppress viscosity in shear flows and only activates it near shocks ($\alpha_{\rm AV}$ ranging from 0.01-2). 
In contrast, StV adopts a constant, maximally dissipative $\alpha_{\rm AV}=2$, producing smoother but overly laminar flows.
As a result, the length of the StV jet lobe reaches $\sim$40-50\% greater than the FID case by $t=100$ Myr. 
We interpret this finding to suggest a high thermal pressure behind the jet head in the StV setup. 
Excess thermal pressure arises from extra thermal energy production due to constant viscosity in solenoidal flow, unlike TDV, where artificial viscosity is applied solely during fluid compression.
At the same time, the high viscosity may effectively clear the ambient gas and carve the low-density region along the jet spine, thereby reducing kinetic energy loss along the jet spine. 
Eventually, the higher viscosity allows more energy to be directly channeled into headward propagation, resulting in narrower and more collimated lobes.

A similar rationale can be applied to the HVisc and LLVisc runs. 
In the HVisc run, the increased artificial viscosity behaves similarly to the StV run on emptying the ambient gas along the jet spine, leading to a deviation from self-similar growth around $t\sim 20 \Myr$, similar to the behavior seen in the StV run. 
Throughout the simulation, the viscosity in the TDV model might decrease when no shock is detected, whereas it is constant in the standard viscosity model. 
This difference results in the sublinear jet lobe expansion in HVisc, unlike the steady linear growth of the jet lobe in the StV run.
In contrast, in the LLVisc run, the initial lower resistance in the ambient medium allows the jet particle to achieve the longest lobe among the viscosity models at the early stage of the jet evolution. 
However, by $t\sim 20 \Myr$, the low viscosity and overmixing between the jet and the ambient hinder the jet head's ability to sweep material in the desired direction, ultimately resulting in the shortest jet lobe among the AV variants at the end of the simulation. 

Among the grid-based jet simulations, the mere example that contains artificial viscosity is by \citet{Krause:2001}. 
Interestingly, although the artificial viscosity suppressed the KH instability in their simulations, it also widened the jet, seemingly the opposite of our results. 
Despite differences in our setups and theirs, artificial viscosity drives bulk motion across the velocity discontinuity in both situations.


\paragraph{Jet duty cycle}
Figure~\ref{fig:DTJetVar-TDKPPP} presents the time evolution of jets with different injection durations: full-duration FID ($t_{\rm jet}=100$ Myr) and truncated cases T50 ($t_{\rm jet}=50$ Myr) and T20 ($t_{\rm jet}=20$ Myr). Despite their early termination, key features, such as jet spines and lobes, persist for several tens of Myr. At $t=21$ Myr, the T20 run still shows a collimated spine; likewise, T50 remains coherent at $t=51$ Myr. However, these features dissipate about 30 Myr after shut-off.

In T20 and T50, lobe growth stalls once momentum injection ceases, but shock front expansion continues. Vertical-velocity and Mach-number slices confirm that the forward flow decays, while backflows and side expansion dominate. Entropy and pressure maps show that cavities remain morphologically distinct even when the axial thrust is gone, consistent with 'ghost bubble' interpretations in cluster observations \citep[e.g.,][]{Churazov2001, Reynolds2005}.

Sudden decreases in lobe length seen in Figure~\ref{fig:Comparison-lobe-length} (at $t\sim65$ Myr for T20 and $\sim80$ Myr for T50) result from our entropy-based lobe identification. When particles near the jet front cool or mix into the ambient medium, their entropy drops below threshold, causing them to fall out of the lobe definition and trigger apparent contraction. This illustrates how tracer definitions can significantly impact quantitative morphology metrics during decaying phases.

\paragraph{Resolution dependence.}
Figures~\ref{fig:Comparison-Resolution} and \ref{fig:Comparison-lobe-length} compare simulations spanning two orders of magnitude in mass resolution ($1.8\times10^5$ to $1.15\times10^7\,M_\odot$). Despite this large range, the global lobe morphology and final extent are relatively robust, with all runs reaching $\sim$300 kpc by $t=98$\,Myr. The lobe length tracks the analytic scaling $L_J\propto t^{3/5}$ within $\sim$10\%, except for early-time deviations in LoD and late-time acceleration in Hi.

Internal structures, however, degrade significantly at lower resolution. The coherent spine and sharp bow shock seen in Hi and FID become poorly defined in LoC and LoD, where numerical viscosity dominates and mixing increases. The vertical velocity and Mach number drop substantially, reflecting reduced thrust efficiency. Entropy maps show broader, rounder lobes, with less directional elongation and greater transverse dispersion.

In grid-based jet simulations, \citet{Rosen:1999} reported an increase in mass entrainment as the spatial resolution was raised. Similarly, \citet{Krause:2001} showed that using a finer grid permits more complex instabilities to form at the jet head, leading to a broader jet and a reduction in jet speed, which contrasts with the behavior seen in our simulation. 
In our case, increasing the resolution in SPH simulations enhances the velocity contrasts, which in turn strengthens the effect of artificial viscosity in these high-resolution SPH runs, contrary to those grid-based simulations.

Importantly, FID shows weaker sensitivity to resolution than StV due to its AV switch: since AV activates only in converging flows, excess diffusion is avoided in marginally resolved regions. In contrast, the StV scheme applies uniform damping, leading to underresolved feedback and exaggerated lobe variations at low resolution.

These findings suggest that while feedback effects can be studied at typical cosmological resolutions, finer details such as metal entrainment, turbulence spectra, or jet stratification require high-resolution runs and controlled AV schemes.

\subsection{Grid-based Counterpart} \label{subsec:grid_sim}

In this section, we present the results from MESH run and compare them with our SPH-based simulations. 

\begin{figure*}
\centering
\includegraphics[width=\textwidth]{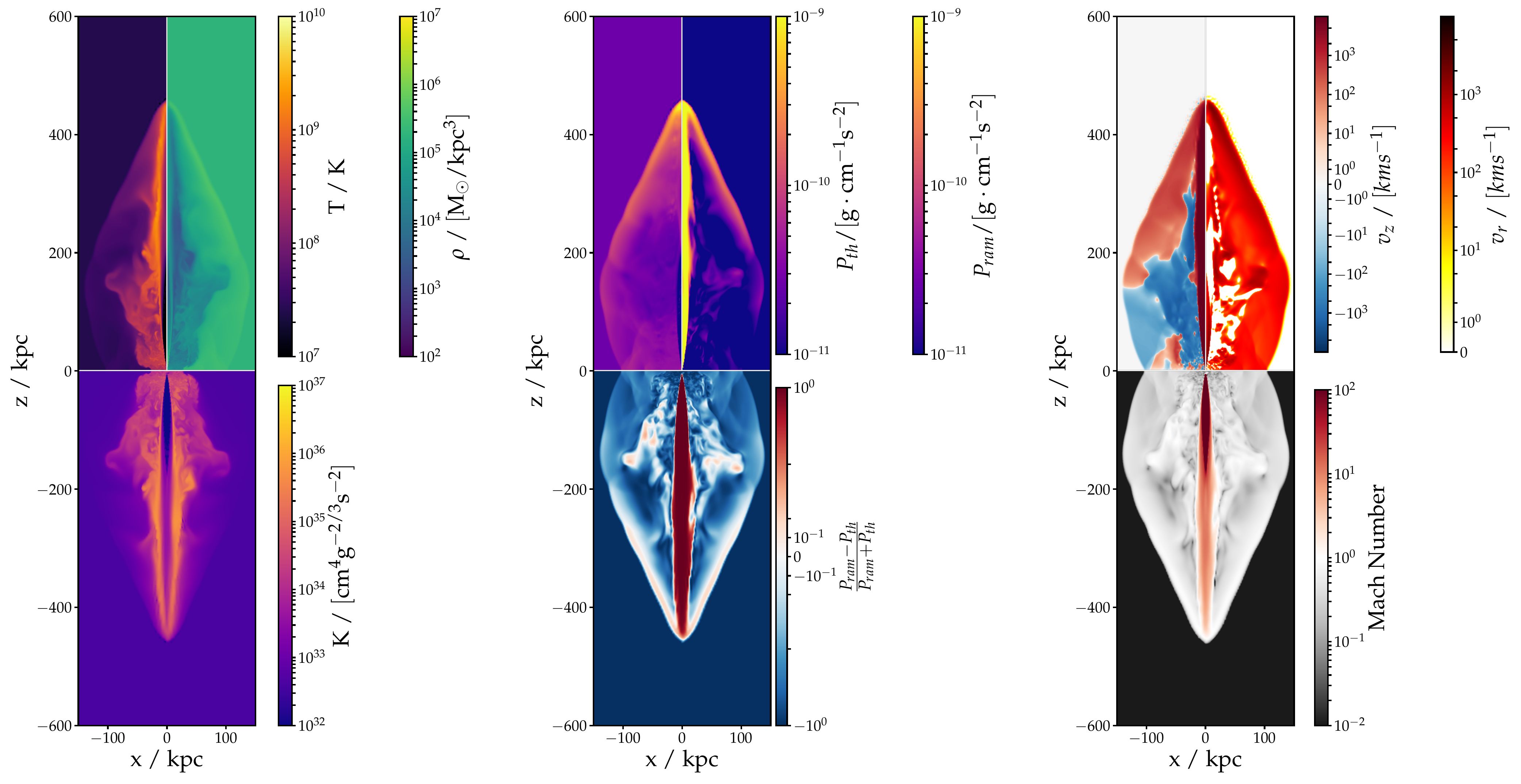}
\caption{\small{
Gas properties at $t = 100\,\mathrm{Myr}$ for the grid-based (MESH) run. The layout is identical to Figure \ref{fig:fid_allmapplot}. 
The jet spine, jet lobe, and jet material can still be identified through discontinuity of thermal and kinetic quantities, even though the morphology and size of the jet lobe are different from the fiducial run.
}}
\label{fig:mesh_allmapplot}
\end{figure*}

The jet morphology is presented in Figure \ref{fig:mesh_allmapplot}. 
Because of the higher spatial resolution of MESH run, we can resolve the structures in much finer detail.  
The jet material, defined as the region with the Mach number $\mathcal{M}>2$, extends to $z\simeq\pm500\,\mathrm{kpc}$. Its overall transverse size is smaller than in the SPH run, but the sharpened jet head follows the same general behavior as in the SPH jet. 
Within the jet material, there is a strongly supersonic segment between the launch point and $z\simeq \pm 200\,\mathrm{kpc}$; beyond this distance, the flow remains supersonic albeit much milder. 
A similar trend is observed in the SPH simulation, but it becomes evident only at higher mass resolution.  
A lobe is still present outside the jet material, but its morphology and dynamics differ from those in the SPH case. 
The lobe is generally narrower, and vortex-like features develop in the middle of the lobe, in contrast to the more inflated and puffy lobe seen in the SPH runs.  
The bow-shock morphology broadly agrees with that of the SPH simulations; the bow-shock front is sharply pointed, indicating a late-time acceleration of the jet head. 
As a result, the overall morphology more closely resembles that of the StV or HVisc runs with enhanced artificial viscosity than that of the fiducial SPH setup.

The evolution of the jet size in MESH run is shown in Figure~\ref{fig:fid_lobesize_evo}. 
For both the jet lobe and the bow shock lengths, MESH run follows the self-similar model and matches the fiducial SPH run to $t=50\,\Myr$; beyond this time, the jet advances more rapidly than predicted by the self-similar solution, consistent with the sharp bow shock visible in Figure \ref{fig:mesh_allmapplot}. 
This late-time acceleration also appears in the fiducial run, but is substantially weaker; only the StV or HVisc runs exhibit a comparable level of acceleration (Figure \ref{fig:Comparison-lobe-length}). 
The measured lobe radius exhibits larger fluctuations relative to the self-similar expectation, which can be attributed to the higher spatial resolution of the mesh grid and the limitations of using 2D slice data for measurements.

In Figure~\ref{fig:fid_energy_evo}, we also present the energy budget of MESH run. 
We find that the evolution of the thermal-kinetic energy partition in MESH run is generally consistent with the fiducial SPH run: the kinetic energy initially dominates but is eventually overtaken by the thermal component. However, in MESH run this transition occurs later at about $t=50\,\Myr$, compared to $t=30\,\Myr$ in the FID run. 
In addition, the thermal-to-kinetic energy ratio in MESH run remains lower than in the FID run at the end of the simulation.

\section{Discussion}
\label{sec:discussion}

In this study, we conducted simulations of AGN jets with an idealized ICM setup. 
The primary focus of our simulations is the convergence of hydrodynamics when varying conditions of jet launching implementations, artificial viscosity configurations, mass resolutions, and jet launching durations. 
To accomplish this, we performed 13 simulation variants in addition to the fiducial one (Table \ref{table:list-of-simulation}). 
In Section \ref{subsec:comp-analytical}, we compare our simulation results with self-similar predictions. 
Section \ref{subsec:comp-SPH} is a comparative analysis of our simulation with \Husko{}, wherein we sought to closely replicate their simulation setup. 
In Section \ref{subsec:comp-MESH}, we compared our SPH simulation with the grid-based counterpart. 
Finally, Section \ref{subsec:comp-observation} explores the connection between our jet-time simulation variants and the Milky Way Fermi bubble.

\subsection{Comparison with analytical solution}\label{subsec:comp-analytical}

In this section, we compare our simulation results with analytical predictions. Figure \ref{fig:fid_lobesize_evo} shows the evolution of the jet lobe in our fiducial simulation alongside the self-similar model of \citet{Kaiser2007}. Overall, the simulation exhibits good agreement with the self-similar theory, although some deviations are present in the detailed growth trend of the lobe.

As shown in Figure \ref{fig:Comparison-lobe-length}, the jet lobe initially grows in a sub-self-similar manner, following $L \propto t^{\alpha}$ with $\alpha < 3/5$ for $0 < t < 40\,\Myr$. Beyond this point, the growth rate increases, and the lobe length eventually catches up with the self-similar prediction by the end of the simulation.

When examining the jet lobe evolution across different artificial viscosity models, we find that all simulations—except for the LLVisc run, which maintains low viscosity throughout—deviate from self-similar growth around $t \sim 50\,\Myr$.

Combining this fact with the evolution of the jet lobe of other variants of artificial viscosity, we find all these simulations, except LLVisc, which has a low viscosity throughout the simulation, deviate from self-similar jet lobe growth around $t\sim 50\Myr$. 

An alternative approach to modeling jet head motion, proposed by \citet{Begelman1989}, balances the jet’s thrust against the ram pressure of the ambient medium. In contrast to the self-similar framework, this model explicitly incorporates the effect of ambient pressure, making it particularly applicable to the later stages of jet evolution. Specifically, it becomes relevant once the jet length surpasses the characteristic scale $L_2$ (Equation~\ref{eq:L2}), beyond which the ambient pressure is no longer negligible. This transition marks the onset of a pressure-confined regime, where the expansion decelerates and the lobe morphology becomes increasingly governed by pressure equilibrium with the surrounding environment.

Assuming that the jet eventually attains a terminal velocity while propagating through a homogeneous ambient medium, one can estimate the lobe’s advance speed by equating the jet thrust with the ram pressure of the surrounding gas (see Appendix~\ref{app:lobe-evolution} for details). This balance yields the following expression for the jet head velocity:
\begin{equation}
\label{Eq:balancing-jet-ram-vel}
    v_\mathrm{H} = \frac{P_\mathrm{j}}{2 \rho_\mathrm{A} v_\mathrm{j}^2 \wsa} \left( \sqrt{1 + \frac{4 \rho_\mathrm{A} \wsa v_\mathrm{j}^3}{P_\mathrm{j}}} - 1 \right),
\end{equation}
where $\rho_\mathrm{A}$ is the ambient gas density, $\wsa$ is the working surface area at the jet head, $v_\mathrm{j}$ is the velocity of the jet material, and $P_\mathrm{j}$ is the jet thrust (momentum flux). This expression captures the transition between the low-thrust and high-thrust regimes and explicitly incorporates the geometrical effect of the jet-ambient interface through $\wsa$.


Unlike the self-similar expression in Equation~\ref{Eq:self-similar-vel}, the lobe propagation speed derived in Equation~\ref{Eq:balancing-jet-ram-vel} is not explicitly time-dependent. Nevertheless, it may still exhibit an implicit time dependence in practice, arising from the gradual evolution of the working surface area, spatial variations in the ambient density, or other environmental inhomogeneities that affect the jet-ambient interaction over time.

As the working surface area $\wsa$ is difficult to measure in our simulation, we try to estimate it from approximate relations with other well-defined quantities. 
One may approximate $\wsa$ with the cross-sectional area of jet head $\pi R_{J,\mathrm{head}}^2$. 
To measure the radius, we first identify the jet material as jet-launched particles with $\mathcal{M}>2$ and measure its length $L_J$. We then define the jet head particles as the jet material located within the outermost $20\%$ of this length.
\begin{equation}
    |z - z_{BH}| > 0.8 L_J
\end{equation}
Finally, the jet head radius $R_{J,\mathrm{head}}^2$ is measured with the method in Section \ref{subsec:measurements}.
This allows us to evaluate the jet head velocity using Equation~\ref{Eq:balancing-jet-ram-vel} with a physically motivated estimation. 

\begin{figure}
\centering
\includegraphics[width=\columnwidth]{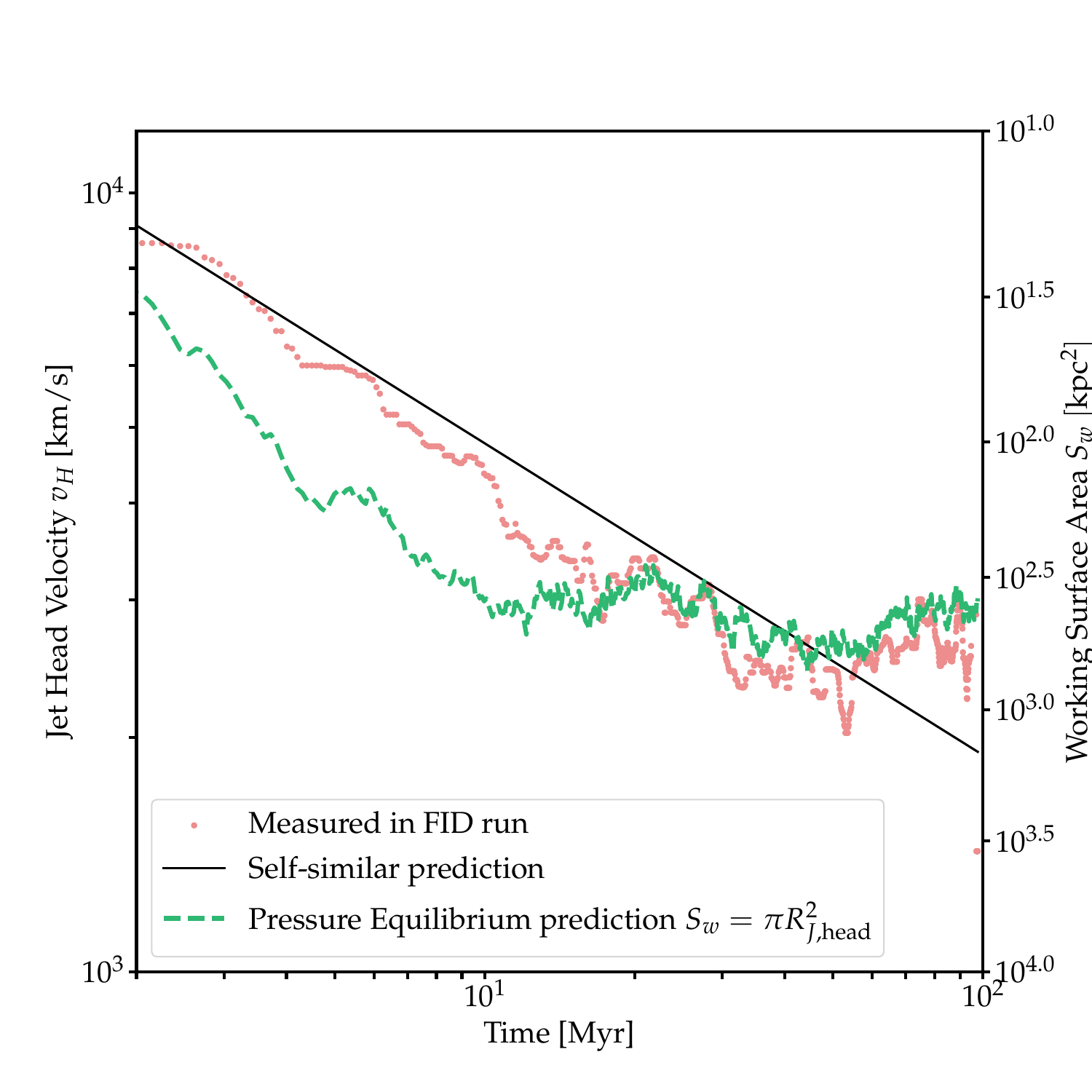}
\caption{\small{Comparison between the measured jet head velocity (bright red dots), the prediction from the self-similar model (solid black line), and the pressure equilibrium model assuming the working surface area equals to the cross-sectional area of the jet head (green dashed line).
A median filter is applied to the measured velocity to suppress noise. 
The right vertical axis indicates the working surface area $S_\mathrm{w}$ corresponding to the jet head velocity, connected via the quadratic relation in Equation~\ref{Eq:balancing-jet-ram-vel}. 
At early times, the measured velocity closely follows the self-similar prediction, while at later times, it transitions toward the pressure equilibrium regime, consistent with a shrinking working surface area.
}}
\label{fig:ComparingTheory}
\end{figure}

Figure~\ref{fig:ComparingTheory} compares the jet-head velocity measured in the FID simulation with the self-similar (Equation \ref{Eq:length-evo-self-similar-final}) and pressure-equilibrium analytic (Equation \ref{Eq:balancing-jet-ram-vel}) predictions. 
During the first $\sim 20\,\Myr$ the simulation tracks the self-similar solution closely.

The jet head velocity drops below the self-similar value at about $t\simeq 20\,\Myr$, but it speeds up again around $t\simeq 50\,\Myr$. This behavior suggests that the working surface is becoming more focused, consistent with the pressure equilibrium expected from the jet head radius.

The sharpening of jet head is presumably due to the reconfinement geometry of the jet material (e.g. \citealt{Komissarov1998}), which is governed by the properties of jet and the external pressure: the jet flow first recollimates and then reconfines, producing a thin, spindle-like morphology along the jet axis. 
In Figures~\ref{fig:fid_temp_dens_evo} and in the animated version of Figure~\ref{fig:fid_allmapplot}, one can clearly see that the jet head exhibits a flat cross section in the middle of the simulation, but becomes reconfined by the end. This interpretation is also consistent with the variation in jet lobe lengths discussed in Section \ref{subsec:compare-diff-jets}, where the choice of viscosity configuration modifies the degree of jet refinement and, in turn, influences the advance speed of the jet head.

\subsection{Comparison with SPH-based Simulations}
\label{subsec:comp-SPH}

In this study, we retained most choices for jet simulation parameters in our fiducial run to remain consistent with the fiducial setup of \Husko{}. Our fiducial simulation reproduces many of the features reported by \Husko{}, although several notable differences emerge.

Most prominently, in terms of jet morphology, we do not observe the pronounced oscillatory re-collimation pattern characteristic of their fiducial self-similar run. Instead, re-collimation in our simulation occurs only once, producing a pencil-shaped jet spine. 
Meanwhile, we notice that the shape of jet spine in our SPH simulation is in good agreement with other well-established grid-based simulation.
We argue that the distinct re-collimation patterns in fact arise from differing approaches to jet launching: \Husko{} appears to inject a swarm of particles simultaneously, producing a recurring sequence of bubbles along the jet axis.
Such a launching scheme may break the steady-state condition and axisymmetry assumptions commonly adopted in the literature (e.g., \citealt{Bodo:2018, Gourgouliatos:2018}).
In our simulation, we adopted a more randomised and smooth launching scheme, which helps the jet remains steady and reaches a good agreement with the grid-based simulation.

Despite this difference in the jet spine, the morphology of the jet lobe and bow shock is broadly consistent with \Husko{}. In both cases, the lobe begins as a quasi-spherical, overpressured bubble that gradually stretches into a slender, cigar-shaped cavity with a sharp head, while its radial growth remains relatively modest (Fig.~\ref{fig:fid_lobesize_evo}).
This evolution reflects a distinct morphological transition, as discussed in Section \ref{subsec:comp-analytical}.
The bow shock outside, characterized by only mild enhancements in temperature and density, preserves an approximately self-similar aspect ratio throughout the evolution in both our simulations and those of \Husko{}.

Quantitatively, we do find a number of differences in the jet evolution measurements (Figures \ref{fig:fid_lobesize_evo} and \ref{fig:fid_energy_evo}) as follows: 

\paragraph{Jet structure sizes}
The evolution of jet structure sizes of our fiducial simulation is compared to \Husko{} in Figure \ref{fig:fid_lobesize_evo}. The difference is mainly due to the measurement at an early time of the simulation. The length of the jet lobe measured in \Husko{} is initially shorter, gradually catching up with the prediction of theory and our measurement around $t<20 \Myr$. 
The radius of the jet lobe of our measurement deviates from the theoretical prediction and \Husko{} during $t<20 \Myr$, which is likely a resolution-related issue, given $\sim 10\,\mathrm{kpc}$ the scale of the jet launching reservoir.

At the same time, we also recall the difference in our lobe identification criterion (Section \ref{subsubsec:measure-jet-lobe}) from
\Husko{}. 
\Husko{} classified lobe material solely by temperature, labeling any gas particle that ever exceeded $5\times10^{8}\,\mathrm{K}$ (their fiducial threshold) as part of the lobe. 
A temperature-only criterion blurs the distinction between genuine lobe gas and shocked ambient gas because both phases can reach temperatures much higher than the ambient value.
In contrast, our entropy-based selection effectively separates the two components by incorporating density information.  
Jet lobe gas is simultaneously hot and underdense, whereas shocked ambient gas is hot but overdense. 
We argue that this combined density--temperature signature provides a cleaner and more physically motivated identification of the lobe, though its performance is limited when the resolution effects play an important role.

\paragraph{Energy partition}
Because our idealised setup disables gravity, cooling, magnetic fields, and cosmic rays, the only energy source is the kinetic power injected by the jet, which is subsequently converted to thermal energy via shocks and shear.  In a perfectly self-similar system, the time derivatives of kinetic and thermal energy should scale in fixed proportion, so that their individual changes ($\Delta E_{\rm kin}$ and $\Delta E_{\rm th}$) remain in a constant ratio, as reported in \Husko{}.  Figure~\ref{fig:fid_energy_evo} shows that this is not the case in our run: kinetic energy dominates up to $t\simeq 20\;\mathrm{Myr}$, after which thermal energy grows more rapidly and overtakes the kinetic component. 
The two curves cross and the thermal energy dominates at the late stage of the simulation, whereas the ratio of thermal and kinetic energy remains roughly 1:2 in the \textsc{SWIFT} simulations of \Husko{}. 
The difference may arise from the choice of SPH formulation, where we adopted the pressure-entropy one while for \Husko{} it was the energy-density scheme. 

At late times, the kinetic energy curve in our simulation gradually flattens toward late times, suggesting a saturation in momentum injection as the jet transitions from a momentum-driven to pressure-driven phase. The bow shock expands more isotropically, with buoyant uplift dominating over bulk axial thrust. This late-stage evolution again underscores the importance of feedback modeling that allows for mode transitions, e.g., from ballistic jets to thermally buoyant bubbles, which can influence the long-term energy coupling to the CGM. 

Overall, although the large-scale morphology agrees with the analytic self-similar theory, the internal energy partition does not follow the constant-ratio behavior found by \Husko{}. 
This highlights the sensitivity of kinetic-to-thermal conversion -- and thus observable X-ray cavity energetics -- to seemingly minor implementation details.

\subsection{Comparison with Grid-based Simulation}
\label{subsec:comp-MESH}

In this study, we investigate the evolution of jets in an idealized environment using the SPH simulation code \gadgetosaka{}, originally developed as a general-purpose tool for cosmological simulations. 
This approach contrasts with many previous works that ran jet simulations using a dedicated mesh-based hydrodynamic simulation code, motivating us to perform a mesh simulation with identical setup. 
In this section, we discuss the comparison between SPH and grid-based simulations.

Following the results presented in Section \ref{subsec:fiducial-jets} and \ref{subsec:grid_sim}, we conclude that our SPH code agrees well with the mesh-based code on jet propagation and jet energy budget, albeit the lobe structures are not well-resolved in SPH runs.
We find that the jet lobe in our SPH run does not exhibit resolved gas motions, such as vortices, that are observed in grid-based simulations. 
This is likely a consequence of the low-density nature of the jet lobe: because SPH is fundamentally mass-weighted, there are too few particles in this region to accurately capture its internal kinematics. 
Despite the limited resolution of the lobe's internal structure in our SPH calculations, the method nonetheless reproduces the jet material properties and length evolution, as well as the morphology of the bow shock, provided that the artificial viscosity parameters are tuned appropriately. 
In addition, the SPH framework can readily incorporate a wide range of additional physics---including gravity, radiative cooling, and sub-grid physics such as star formation and cosmic rays, making it particularly well suited for jet simulations in more realistic setups and better aligned with the requirements of galactic astrophysics studies.

A complementary perspective is offered by the clumpy-ISM study of \citet{Dutta2024}. Using analytic arguments calibrated with 3-D grid-based simulations, they show that once the effective density ratio between clouds and diffuse gas exceeds $\lambda\approx 20$, the jet head stalls, the bow shock inflates quasi-isotropically, and most of the mechanical power dissipates locally (their Eq. 27 and Figure 3).  
Our idealised runs occupy the opposite extreme, $\lambda\simeq 1$, where the jet remains momentum-driven and follows the self-similar $L_J\propto t^{3/5}$ growth until late-time turbulence steepens the head advance.  The two studies therefore bracket the range of ambient inhomogeneity likely encountered by real AGN.  Incorporating a clumpy CGM into \gadgetosaka{} --- while retaining the low-viscosity, reservoir-based launch scheme validated here --- will allow a direct SPH test of \citet{Dutta2024}’s dissipation criterion and of how numerical viscosity modulates the anisotropic vs. isotropic transition.

\subsection{Jet remnant evolution}\label{subsec:comp-observation}

Our truncated-jet runs (T20, T50) demonstrate that an over-pressurised cavity remains visible for tens of Myr after energy injection ceases, with the bow shock continuing to expand buoyantly.  Rescaled to Sgr A*, a $P_{\rm j}\sim 10^{41}$\,erg\,s$^{-1}$ jet active for $20\!-\!30$\,Myr would leave a $10^{56}$\,erg thermal bubble comparable to the eROSITA shell, while a younger, more energetic burst could account for the sharper, higher-luminosity Fermi edges. 
We conclude the aspect ratio of near-spherical eROSITA bubble morphology can be achieved with a shorter but more energetic burst. 
At the same time, the variation on the jet properties (e.g., lighter collimated jet in \citealt{Zanni:2003}; higher external Mach number and larger opening angle in \citealt{Krause:2012}) or the environment setup (e.g., a steeper density profile or clumpy halo in \citealt{Dutta2024}) may also contribute to the morphology of the bubble.
Incorporating complexities—together with cosmic-ray transport—will be the next step in linking idealised jet simulations to multi-wavelength observations of Galactic outflows.

\section{Conclusions}
\label{sec:conclusions}

We have performed a suite of hydrodynamic simulations of kinetic AGN jets using \gadgetosaka{} SPH code over a duration of $\sim$100\,Myrs, following the work by \citet{Husko2023a}. 
Our fiducial simulation not only successfully reproduces the principal characteristics predicted by classical analytical models, but also exhibits good agreement with grid-based simulations at limited mass resolution, while retaining sufficient flexibility to incorporate additional physical processes.
In a uniform ambient medium, the jet lobe length follows the expected self-similar scaling $L_{\rm j} \propto t^{3/5}$ to within 10\% for most of the evolution. Deviations appear at late times when ambient pressure becomes significant, at which point the jet-head velocity transitions smoothly from the self-similar regime to a pressure-equilibrium regime. This transition is well captured by the momentum-balance model derived in Appendix~\ref{app:lobe-evolution}.
To further characterize the jet structure, we decomposed it into distinct components and examined their thermodynamic evolution in the $\rho$--$T$ phase diagram.

We have also explored the sensitivity of jet morphology to various numerical schemes. Switching from a pre-allocated particle reservoir to on-the-fly spawning at a finite distance shortens the final lobe length by 15--20\%, due to delayed and more dispersed momentum injection. Nonetheless, the large-scale morphology remains broadly consistent. Artificial viscosity prescriptions play a more significant role: adopting a constant, maximally dissipative viscosity or a high-ceiling time-dependent model enhances jet collimation and extends the lobe by 40--50\% relative to the fiducial model. In contrast, a low-viscosity ceiling leads to early mixing and deviation from self-similar growth as early as $t \sim 50\,\Myr$.

When jet power is switched off at $t=20$--$50\,\Myr$, the resulting fossil lobes decelerate and contract, as high-entropy material gradually mixes into the ambient gas. These results highlight how tracer definitions and feedback history affect late-time lobe morphology. We also tested convergence with mass resolution: varying the particle mass by two orders of magnitude yields final lobe lengths within 10\% of the analytic expectation. However, fine structural features—such as bow-shock sharpness, backflows, and metal entrainment—are significantly degraded at lower resolution or with constant-$\alpha$ viscosity, demonstrating the importance of high resolution and adaptive AV schemes for capturing CGM/ICM-relevant physics.

Looking forward, our study lays the foundation for several promising directions. Embedding jets in stratified cluster atmospheres with gravity, cooling, magnetic fields, and cosmic rays will be essential for capturing realistic buoyancy, mixing, and non-thermal pressure support. Coupling this feedback model to self-regulated black hole growth in cosmological simulations will enable studies of jet duty cycles and their broader impact on galaxy populations. Synthetic X-ray and radio maps informed by our detailed lobe structure could provide valuable comparisons to upcoming observations from \textit{Athena}, SKA, and other facilities. Finally, the implementation of Lagrangian tracers, subgrid turbulence models, or differentiable GPU-accelerated hydrodynamics will help reduce numerical uncertainties and enable field-level inference of jet parameters.

In summary, our results demonstrate that \gadgetosaka{} is capable of reproducing analytic jet scalings, identifying key numerical dependencies, and serving as a robust platform for next-generation studies of AGN feedback in galaxy formation and evolution.

\section*{Acknowledgments}
We thank the anonymous reviewer for the precious suggestions on the improvement of this manuscript.
We are grateful to Filip Hu\v{s}ko, Cedric Lacey for the discussion about their work on jet simulation.  
C.D. thanks Enrico Garaldi for a useful discussion on the artificial viscosity schemes. 
KN is grateful to Volker Springel for providing the original version of {\sc GADGET-4}, on which the \gadgetosaka{} code is based. 
Some of the numerical computations for this study were carried out on the computational cluster \texttt{idark} at Kavli IPMU, the Cray XC50 at the Center for Computational Astrophysics, National Astronomical Observatory of Japan, the {\sc SQUID} at the Cybermedia Center, Osaka University as part of the HPCI system Research Project (hp240141, hp250119),
and the HOKUSAI BigWaterfall2(HBW2) at RIKEN. 
This research was supported by FoPM, WINGS Program, the University of Tokyo.
This work is supported in part by the MEXT/JSPS KAKENHI grant number 20H00180, 22K21349, 24H00002, and 24H00241 (K.N.)
and 25H00675 (A.M.).  
K.N. acknowledges the support from the Kavli IPMU, World Premier Research Center Initiative (WPI), UTIAS, the University of Tokyo. 
R.C. acknowledges in part financial support from the start-up funding of Zhejiang University and Zhejiang provincial top level research support program.

\section*{Data Availability}

The snapshots of \gadgetosaka{} jet simulation, as well as the analysed data products, will be shared upon request to the authors. The animated visualisation of the simulation is available on YouTube: \url{https://www.youtube.com/@KNastro-cosmos}.



\bibliographystyle{mnras}
\bibliography{master}



\appendix

\section{Lobe head motion}
\label{app:lobe-evolution}

We model the advance of the jet (lobe) head into the ambient medium by balancing the jet thrust against the ambient ram (and, when needed, thermal) pressure at the working surface \citep[e.g.,][]{Bourne2017, Begelman1989}. 
Considering one side of the jet, Newton's law across the head gives
\begin{equation}
\sum \boldsymbol{F}_\mathrm{H}
= \dot m_j\,(\boldsymbol{v}_j-\boldsymbol{v}_\mathrm{H})
  + \rho_A\,\bar S_w\,(\boldsymbol{v}_\mathrm{A} - \boldsymbol{v}_\mathrm{H})\,|\boldsymbol{v}_\mathrm{A} - \boldsymbol{v}_\mathrm{H}|
= M_\mathrm{H}\,\boldsymbol{a}_\mathrm{H} ,
\tag{A1}
\end{equation}
where $\dot m_j$ is the jet mass outflow rate, and $\bar S_w$ is the \emph{effective} working surface area through which the head couples to the ambient gas. We parametrize
\begin{equation}
\bar S_w \equiv \xi\, S_w , \qquad 0<\xi\le 1 ,
\tag{A2}
\end{equation}
with $S_w$ the geometric area and $\xi$ a dimensionless efficiency factor that accounts for obliquity/porosity of the working surface.

Note that in this analysis, only one side of the jet (or lobe) is considered, i.e. $\dot{m}_\mathrm{j} = \dot{M}_\mathrm{j} / 2$. 
$\vec{v}_\mathrm{H}$ is the velocity of the lobe head, $\vec{v}_\mathrm{A}$ and $\rho_\mathrm{A}$ are the velocity and the density of the ambient gas respectively. 
Upon being ejected from the black hole, jets undergo minimal deceleration while travelling through the low-density environment within the jet lobe cavity. However, once they encounter the ambient medium, the jet rapidly decelerates, reaching a terminal velocity. 

Assuming a static ambient medium ($\boldsymbol{v}_\mathrm{A}=\boldsymbol{0}$), a fixed jet direction, and a quasi-steady head ($\boldsymbol{a}_\mathrm{H}=\boldsymbol{0}$), the axial momentum balance reduces to
\begin{equation}
\dot m_j\,(v_j - v_\mathrm{H}) - \rho_\mathrm{A}\,\bar S_w\,v_\mathrm{H}^2 = 0 .
\tag{A3}
\end{equation}
Define $v_0 \equiv \dot m_j/(\rho_\mathrm{A} \bar S_w)$. When $v_j \gg v_\mathrm{H}$ (the classical limit; \citealt{Bourne2017,Begelman1989}), Eq.~(A3) gives
\begin{equation}
v_\mathrm{H} \simeq \sqrt{v_0\,v_j}, \qquad v_0=\frac{\dot m_j}{\rho_\mathrm{A} \bar S_w}.
\tag{A4}
\end{equation}
Without taking $v_j\!\gg\!v_\mathrm{H}$, the exact solution of Eq.~(A3) is the positive root of the quadratic,
\begin{equation}
v_\mathrm{H}=\frac{v_0}{2}\!\left(\sqrt{1+\frac{4\,v_j}{v_0}}-1\right),
\tag{A5}
\end{equation}
which satisfies $0 < v_\mathrm{H} < v_j$.


To include thermal pressure, work in the head frame and balance the normal stresses (e.g., \citealt{Matzner2003}),
\begin{equation}
\rho_j (v_j - v_\mathrm{H})^2 + p_j \;=\; \rho_\mathrm{A} v_\mathrm{H}^2 + p_\mathrm{A} .
\tag{A6}
\end{equation}
Relating the jet density at the working surface to the mass flux,
\begin{equation}
\rho_j \;=\; \frac{\dot m_j}{S_w\,(v_j - v_\mathrm{H})} ,
\tag{A7}
\end{equation}
and neglecting thermal pressures ($p_j,p_A\to 0$) recovers the standard thrust-ram result \citep{Marti1994},
\begin{equation}
\frac{v_\mathrm{H}}{v_j}=\frac{\sqrt{\eta_\rho}}{1+\sqrt{\eta_\rho}},
\qquad \eta_\rho \equiv \frac{\rho_j}{\rho_\mathrm{A}}.
\tag{A8}
\end{equation}


When thermal pressure differences matter, define
\[
\beta \equiv \frac{p_j - p_A}{\rho_\mathrm{A} v_j^2},
\qquad
\eta_\rho \equiv \frac{\rho_j}{\rho_\mathrm{A}}.
\]
Combining Eq.~(A6) with these definitions yields a quadratic for $x\equiv v_H/v_j$ whose physical root ($0<x<1$ for light jets) is
\begin{equation}
\frac{v_\mathrm{H}}{v_j}
= \frac{\sqrt{\eta_\rho + \beta(1-\eta_\rho)} - \eta_\rho}{\,1-\eta_\rho\,}
\tag{A9}
\end{equation}
which reduces exactly to Eq.~(A8) for $\beta\to 0$.

If one prefers to normalize the pressure jump by the \emph{jet} ram pressure, let
\[
\Pi \equiv \frac{p_j - p_\mathrm{A}}{\rho_j v_j^2}=\frac{\beta}{\eta_\rho},
\]
and write Eq.~(A9) equivalently as
\begin{equation}
\frac{v_\mathrm{H}}{v_j}
= \frac{\sqrt{\eta_\rho\!\left[1+\Pi(1-\eta_\rho)\right]} - \eta_\rho}{\,1-\eta_\rho\,}.
\tag{A10}
\end{equation}

\paragraph{Useful limits.}
(i) $\beta\to 0$: $v_\mathrm{H}/v_j\to \sqrt{\eta_\rho}/(1+\sqrt{\eta_\rho})$ (Eq.~A8).
(ii) $\eta_\rho\to 0$ at fixed $\beta$: $v_\mathrm{H}/v_j\to \sqrt{\beta}$, i.e. the head speed is set by jet overpressure against ambient inertia.
(iii) Large overpressure $\Pi\gg 1$: $v_\mathrm{H}/v_j\simeq \sqrt{\eta_\rho\,\Pi}$ until $v_\mathrm{H}$ approaches $v_j$.


\bsp	
\label{lastpage}
\end{document}